\documentclass{pasa}

\newcommand{\apjl}{ApJ}
\newcommand{\apjs}{ApJS}
\newcommand{\aap}{A\&A}

\newcommand{\Ang}{\mbox{ \AA}}     

\newcommand{\avg}[1]{\ensuremath{\langle #1 \rangle}}
\newcommand{\bma}{\begin{math}}
\newcommand{\ema}{\end{math}}
\newcommand{\beq}{\begin{equation}}
\newcommand{\eeq}{\end{equation}}
\newcommand{\beqa}{\begin{eqnarray}}
\newcommand{\eeqa}{\end{eqnarray}}
\newcommand{\bc}{\begin{center}}
\newcommand{\ec}{\end{center}} 
\newcommand{\bit}{\begin{itemize}}
\newcommand{\eit}{\end{itemize}}

\newcommand{\Lya}{Ly$\alpha$ }
\newcommand{\Lyb}{Ly$\beta$ }
\newcommand{\Lyg}{Ly$\gamma$ }
\newcommand{\hi}{H~{\sc i}}
\newcommand{\oi}{O~{\sc i}}
\newcommand{\cii}{C~{\sc ii}}
\newcommand{\civ}{C~{\sc iv}}
\newcommand{\siii}{Si~{\sc ii}}
\newcommand{\siiv}{Si~{\sc iv}}

\newcommand{\mgii}{Mg~{\sc ii}}
\newcommand{\feii}{Fe~{\sc ii}}

\def\HI{\hbox{H~$\rm \scriptstyle I\ $}}

\def\HeII{\hbox{He~$\rm \scriptstyle II\ $}}

\font\BFd=cmmib10
\font\BFt=cmmib10
\font\BFs=cmmib10 scaled 700
\font\BFss=cmmib10 scaled 500

\def\bbox#1{%
\relax\ifmmode
\mathchoice
{{\hbox{\BFd #1}}}
{{\hbox{\BFt #1}}}
{{\hbox{\BFs #1}}}
{{\hbox{\BFss #1}}}
\else \mbox{#1} \fi }

\title[Reionization: quasar absorption lines]{Reionization and high-redshift galaxies: the view from quasar absorption lines}
\author[G.D. Becker, J.S. Bolton \& A. Lidz]{George D. Becker$^{1,2}$\thanks{george.becker@ucr.edu}, James S. Bolton$^3$\thanks{james.bolton@nottingham.ac.uk}, Adam Lidz$^4$\thanks{alidz@sas.upenn.edu}\\
\affil{$^1$Space Telescope Science Institute, 3700 San Martin Drive, Baltimore, MD 21218, USA}
\affil{$^2$Department of Physics \& Astronomy, University of California, Riverside, 900 University Avenue, Riverside, CA 92521, USA}
\affil{$^3$School of Physics and Astronomy, University of Nottingham, University Park, Nottingham NG7 2RD, UK}
\affil{$^4$Department of Physics and Astronomy, University of Pennsylvania, 209 South 33rd Street, Philadelphia, PA 19104, USA}}

\jid{PASA}
\doi{10.1017/pas.\the\year.xxx}
\jyear{\the\year}

\usepackage[authoryear]{natbib}
\bibpunct{(}{)}{;}{a}{}{,}
\usepackage{hyperref}

\begin{document}

\begin{abstract}

Determining when and how the first galaxies reionized the
intergalactic medium (IGM) promises to shed light on both the nature
of the first objects and the cosmic history of baryons.  Towards this
goal, quasar absorption lines play a unique role by probing the
properties of diffuse gas on galactic and intergalactic scales.  In
this review we examine the multiple ways in which absorption lines
trace the connection between galaxies and the IGM near the reionization epoch.  We first describe how the
\Lya forest is used to determine the intensity of the ionizing
ultraviolet background and the global ionizing emissivity budget.
Critically, these measurements reflect the escaping ionizing radiation
from all galaxies, including those too faint to detect directly.  We
then discuss insights from metal absorption lines into
reionization-era galaxies and their surroundings.  Current
observations suggest a buildup of metals in the circumgalactic
environments of galaxies over $z \sim 6$ to 5, although changes in ionization will also affect the evolution of metal line properties.  A substantial
fraction of metal absorbers at these redshifts may trace relatively low-mass galaxies.  Finally, we
review constraints from the \Lya forest and quasar near zones on the
timing of reionization.  Along with other probes of the high-redshift
Universe, absorption line data are consistent with a relatively late
end to reionization ($5.5 \lesssim z \lesssim 7$); however the
constraints are still fairly weak.  Significant progress is expected
to come through improved analysis techniques, increases in the number
of known high-redshift quasars from optical and infrared sky surveys,
large gains in sensitivity from next-generation observing facilities,
and synergies with other probes of the reionization era.

\end{abstract}

\begin{keywords}
quasars: absorption lines -- dark ages, reionization -- intergalactic medium -- galaxies: high-redshift -- galaxies: evolution 
\end{keywords}

\maketitle

\section{INTRODUCTION}
\label{sec:intro}

The period over which the first sources ionized the hydrogen in the
intergalactic medium (IGM) and ended the preceding cosmic ``dark
ages'' is the Epoch of Reionization (EoR), and determining exactly
when and how reionization happened is a key goal for observational and
theoretical cosmology \citep{LoebFurlanetto2013}.  Understanding
reionization is important for several reasons. First, it is
a major event in our cosmic history that impacted on almost every
baryon in the Universe; a better understanding of the EoR is thus
vital for developing a complete picture of the Universe's history.
Second, detailed measurements of IGM properties during reionization
will strongly inform models of the first luminous sources and high-redshift structure formation, addressing a broad range of open
questions regarding the nature of the early Universe.  Third, these
early generations of ionizing sources influenced the formation of
subsequent galaxy populations.  Finally, uncertainties in reionization
physics lead to ``nuisance parameters'' that may limit our ability to
extract cosmological parameters from cosmic microwave background (CMB)
and \Lya forest data sets, amongst others.

Although some aspects of reionization may be understood by studying
high-redshift galaxy populations directly, IGM measurements play an
important complementary role.  After all, it is fundamentally the
interplay between the ionizing sources and the surrounding
intergalactic gas that determines the nature of reionization; the
properties of the intergalactic gas, including its ionization state,
temperature, and metallicity, generally depend on the {\em collective
  impact of all of the luminous sources} while galaxy surveys
typically detect only bright sources that lie above survey flux
detection limits.

In this context, it has been half a century since
\citet{GunnPeterson1965} drew attention to the lack of prominent \Lya
absorption troughs in the spectra of the -- then newly discovered --
quasars.  The absence of strong absorption revealed that there was
very little intervening neutral hydrogen in intergalactic space, all
the way out to the highest-redshift object observed at the time,
quasar 3C 9 at $z=2.01$ \citep{Schmidt1965}.  In the intervening fifty
years, there has been tremendous progress in the study of the IGM
using quasar absorption lines, and we now have detailed constraints on
many of properties of the IGM and the EoR which extend to the
current highest-redshift quasar, ULAS J1120$+$0641 at $z =7.085$
\citep{Mortlock2011}.  The aim of the present article is to review
these constraints, examine their implications, and consider the
prospects for improving them in the future.

\begin{figure*}
\begin{center}
\includegraphics[width=\textwidth]{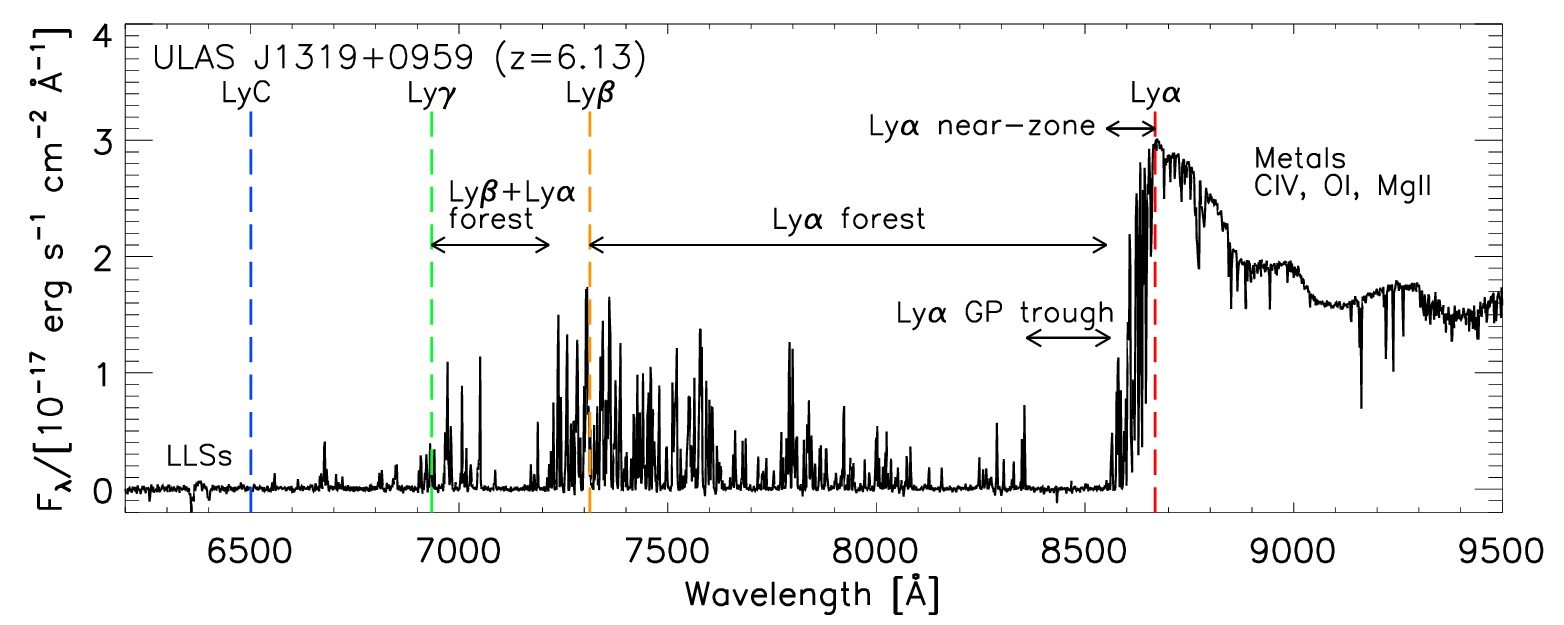}
\vspace{-0.7cm}
\caption{A high signal-to-noise spectrum of the quasar ULAS J1319+0959
  at $z=6.13$ from \citet{Becker2015}, obtained with the X-Shooter
  spectrograph on the Very Large Telescope (VLT). The spectrum has
  been rebinned to $1.5\rm\,\AA$ per pixel for presentation purposes.
  This illustrates many of the features reviewed here -- see the text
  in \S \ref{sec:intro} for a description.}
\label{fig:qso_examp}
\end{center}
\end{figure*}

In \S \ref{sec:uvb} we review the properties of the ultraviolet
background inferred from the post-reionization \Lya forest at $z \leq
6$.  We compare these measurements to the number of ionizing photons
expected from star-forming galaxies and quasars, and assess what these
data imply for the sources likely responsible for reionizing the
IGM. In \S \ref{sec:metals} we review current observations of IGM
metal line abundances at $z>5$, address whether the known galaxy
population approaching and during the EoR can account for the observed
metal enrichment, and consider the implications of metal line
populations for high-redshift galaxy formation.  Direct constraints on
the reionization history using quasar absorption line data are then
described in \S \ref{sec:reion_mods}.  We also briefly compare these
data with other, complementary probes of reionization.  Finally, in \S
\ref{sec:conclude} we conclude with a discussion of future prospects
for exploring the EoR with quasar absorption lines.

For further orientation, Fig. \ref{fig:qso_examp} provides an example
of a $z \gtrsim 6$ quasar spectrum in the observed-frame; this
illustrates key spectral features used to infer IGM properties
approaching the EoR.  Redward of the \Lya emission from the quasar
(red dashed line), one can identify a series of metal absorption lines
(\S \ref{sec:metals}). Close to the quasar redshift lies the \Lya
proximity or near-zone, where the quasar contributes significantly to
ionizing the hydrogen in its vicinity (\S \ref{sec:prox}). Next,
moving to shorter wavelengths, is the \Lya absorption forest
from intervening neutral hydrogen in the cosmic web (\S
\ref{sec:uvb}). This $z \gtrsim 6$ spectrum also shows a complete
Gunn-Peterson absorption trough (\S \ref{sec:GP}) above $8400 \Ang$
(from hydrogen at $z \gtrsim 5.9$ absorbing in the \Lya line) that
continues until the near-zone region.  Between the green and orange
dashed lines, which mark the wavelengths of the \Lyb and \Lyg
transitions at the quasar systemic redshift, lies the \Lyb
forest.  In this region of the spectrum, high-redshift gas absorbs in
the \Lyb line and at lower redshift, foreground gas absorbs in \Lya
(\S \ref{sec:higher_order}).  At even shorter wavelengths, overlapping
higher-order Lyman series transitions occur.  Finally, below the line
marked ``LyC'' there is continuum absorption from neutral hydrogen:
photons at these wavelengths -- with rest frame wavelength $\lambda
\leq 912 \Ang$ -- are energetic enough to photo-ionize hydrogen atoms.
In lower-redshift quasar spectra where there is less overall
absorption, Lyman limit systems (LLSs) -- absorbers that have an
optical depth of unity to photons at the hydrogen photo-ionization
edge -- can be identified here. LLSs, along with cumulative absorption
from lower-column density absorbers, set the mean free path to
ionizing photons in the IGM (\S \ref{sec:mfp}).

\section{THE UV BACKGROUND} \label{sec:uvb}

The ultraviolet background (UVB) is a key probe of the sources of
hydrogen ionizing photons ($E\geq 13.6\rm\,eV$) in the
post-reionization era at $z<6$; its intensity and spectral shape
provides a complete census of ionizing photon production and its
evolution with redshift
\citep{HaardtMadau1996,HaardtMadau2012,FaucherGiguere2009}.  One of
the primary observational techniques used to probe the UVB is quasar
absorption line spectrosopy.  The \Lya forest -- the observable
manifestation of the intergalactic neutral hydrogen that traces the
cosmic web of large-scale structure \citep[see
  e.g.][]{Rauch1998,Meiksin2009} -- is particularly important in this
regard.  In this section we discuss the theoretical and observational
framework on which UVB measurements using the \Lya forest are based,
and examine the implications of these data for the properties of high-redshift galaxies during the reionization era.

\subsection{The Gunn-Peterson effect} \label{sec:GP}

It is instructive to first review the relationship between the neutral
hydrogen fraction of the IGM and the intergalactic \Lya opacity.
\cite{GunnPeterson1965} demonstrated that quasar spectra are a useful
probe of the intergalactic neutral hydrogen density.  If the
intervening IGM along a quasar line-of-sight contains sufficient
neutral hydrogen, the transmitted flux blueward of a quasar's \Lya
emission line (at rest-frame wavelength $\lambda_{\alpha}=1215.67\rm
\AA$) will be completely attenuated, producing a saturated absorption
trough.

The \cite{GunnPeterson1965} argument is as follows.  Consider light
emitted by a quasar at redshift $z_{\rm q}$ passing through a uniform
IGM with proper neutral hydrogen number density $n_{\rm HI}(z)$.  This
light is observed at redshift $z<z_{\rm q}$ with frequency
$\nu=\nu_{\alpha}/(1+z)$, such that the emitted photons have
redshifted into \Lya resonance with the local IGM.  The total optical
depth along the line-of-sight is then
\begin{equation} \tau_{\rm GP}^{\alpha} = \int_{0}^{z_{\rm q}}
  \sigma_{\rm s}[\nu(1+z)]n_{\rm HI}(z) \frac{dl}{dz}
  dz, \end{equation} where $dl/dz=-c/[H(z)(1+z)]$ is the proper line
element.  Ignoring line broadening effects, the effective scattering
cross-section, $\sigma_{\rm s}$, may be approximated by a Dirac delta
function peaked at $\nu_{\alpha}$
\[\sigma_{\rm s}[\nu(1+z)] = \sigma_{\alpha}\nu_{\alpha}\delta[\nu(1+z)-\nu_{\alpha}], \]
where $\sigma_{\alpha}=4.48\times 10^{-18}\rm\,cm^{2}$ is the \Lya
cross-section.  Changing the variable of integration to $\nu$, where
$dl/d\nu=\lambda_{\alpha}(1+z)/H(z)$ and noting that $\tau_{\rm
  GP}^{\alpha}=0$ when $\nu \leq \nu_{\alpha}/(1+z)$ then yields
\begin{equation} \tau_{\rm GP}^{\alpha}\simeq \frac{\sigma_{\alpha}c  n_{\rm HI}(z)}{H_{0}\Omega_{\rm m}^{1/2}(1+z)^{3/2}},  \label{eq:GP}  \end{equation}
using the high-redshift ($z\geq 2$) approximation for the Hubble
parameter, $H(z)\simeq H_{0}\Omega_{\rm m}^{1/2}(1+z)^{3/2}$.
Identifying $\avg{x_{\rm HI}}=n_{\rm HI}/{\bar n}_{\rm H}$ as the
average neutral hydrogen fraction, the Gunn-Peterson optical depth at
the background density, ${\bar n}_{\rm H}=\rho_{\rm crit}\Omega_{\rm
  b}(1-Y)(1+z)^{3}/m_{\rm H}$, is then
\begin{align} \tau_{\rm GP}^{\alpha} &\simeq 2.3 \times 10^{5} \, \avg{x_{\rm HI}} \left(\frac{\Omega_{\rm b}h^{2}}{0.022}\right)\left(\frac{\Omega_{\rm m}h^{2}}{0.142}\right)^{-1/2} \nonumber \\
&\times
  \left(\frac{1-Y}{0.76}\right)\left(\frac{1+z}{5}\right)^{3/2}, \label{eq:evalGP} \end{align}
where $Y$ is the primordial helium fraction by mass.  

The transmittance shortward of a quasar's \Lya emission line is just
$e^{-\tau_{\rm GP}^{\alpha}}$.  Consequently, even for a modest
neutral fraction of $\avg{x_{\rm HI}}\sim 10^{-4.5}$ the Gunn-Peterson
optical depth is fully saturated (i.e. $e^{-\tau_{\rm
    GP}^{\alpha}}\approx0$).  Observationally, the decline in the observed
\Lya opacity and especially the absence of \Lya troughs in quasar spectra at
$z<5.5$ indicates the volume-weighted neutral hydrogen fraction in the
IGM is very small by this redshift
(\citealt{Becker2001,Djorgovski2001,Songaila2004,Fan2006}, see also
\citealt{McGreer2015}) -- we shall return to this point in
\S\ref{sec:meanf_a}.

\subsection{The \Lya forest opacity and the metagalactic hydrogen ionization rate} \label{sec:FGPA}

The \cite{GunnPeterson1965} argument implies the IGM is highly ionized
along observed quasar sight-lines at $z<5.5$.  However, it does not
directly relate the observed \Lya opacity to the quantity of interest
here -- the intensity of the UVB.  To progress further, we must
recognise that intergalactic \Lya absorption arises not from a uniform
medium, but the continuous, fluctuating distribution of baryons which
forms through hierarchical structure formation within cold dark matter
models.  Estimates of the ionizing photon production by sources in the
early Universe rely on the resultant relationship between the opacity
of the \Lya ``forest'' of absorption lines and the intensity of the
UVB.

We may consider the relationship between the \Lya forest opacity and
the UVB as follows.  The UVB specific intensity at redshift $z_{0}$
and frequency $\nu_{0}$, is given by
(e.g. \citealt{HaardtMadau1996,FaucherGiguere2008,BeckerBolton2013})
\begin{equation} J(\nu_{0},z_{0}) = \frac{1}{4\pi}\int_{z_{0}}^{\infty}dz\frac{dl}{dz}\frac{(1+z_{0})^{3}}{(1+z)^{3}}\epsilon(\nu,z)e^{-{\bar \tau}(\nu_{0},z_{0},z)}. \label{eq:Jnu} \end{equation}
This expression is obtained by solving the cosmological radiative
transfer equation, where $\epsilon(\nu,z)$ is the proper specific
emissivity, $\nu=\nu_{0}(1+z)/(1+z_{0})$ and ${\bar
  \tau}(\nu_{0},z_{0},z)$ is the intervening effective optical depth
for photons with frequency $\nu_{0}$ at redshift $z_{0}$ that were
emitted at redshift $z$ (see Eq. \ref{eq:tau_ll} later).  The number
of hydrogen atoms photo-ionized per unit time, $\Gamma_{\rm HI}$, is
then related to the specific intensity of the UVB by
\begin{equation}  \Gamma_{\rm HI}(z) = \int_{\nu_{\rm LL}}^{\infty} \frac{4\pi J(\nu,z)}{h_{\rm P}\nu} \sigma_{\rm HI}(\nu)d\nu \simeq \frac{4\pi J_{\rm LL}\sigma_{\rm LL}}{h_{\rm P}(\alpha_{\rm bg}+3)}, \label{eq:Gamma} \end{equation} 
where $\sigma_{\rm HI}$ is the \HI photo-ionization cross-section
\citep[see e.g.][]{VernerFerland1996b}, $h_{\rm P}$ is Planck's
constant and $\nu_{\rm LL}$ is the photon frequency at the Lyman
limit.  The final step makes the simplifying assumption of a power-law
UVB spectrum, $J(\nu)=J_{\rm LL}(\nu/\nu_{\rm LL})^{-\alpha_{\rm
    bg}}$, and approximates\footnote{A scaling of $\sigma_{\rm HI}
  \propto \nu^{-2.75}$ is a better fit to the true frequency
  dependence at energies close to 1 Ryd
  \citep[e.g.][]{osterbrock2006}; however, we adopt the commonly used
  approximation $\sigma_{\rm HI} \propto \nu^{-3}$ for simplicity.}
the photo-ionization cross-section as $\sigma_{\rm HI}\simeq
\sigma_{\rm LL}(\nu/\nu_{\rm LL})^{-3}$, where $\sigma_{\rm
  LL}=6.35\times10^{-18}\rm\,cm^{2}$.

The photo-ionization rate can be related to the neutral hydrogen
fraction in the IGM.  Following reionization, hydrogen in the low-density IGM is in ionization equilibrium with the UVB.  If ignoring
collisional ionizations (appropriate for the low temperatures,
$T<10^{5}\rm\,K$ associated with gas in the \Lya forest),
\begin{equation} n_{\rm HI}\Gamma_{\rm HI}=n_{\rm e}n_{\rm HII}\alpha_{\rm HII}(T). \label{eq:eq} \end{equation}
Here $n_{\rm HI}$, $n_{\rm HII}$ and $n_{\rm e}$ are the number
densities of neutral hydrogen, ionized hydrogen and free electrons,
and $\alpha_{\rm HII}(T)$ is the temperature dependent radiative
recombination coefficient.  We adopt the case-A recombination
  rate $\alpha_{\rm HII}(T)=4.063\times
10^{-13}(T/10^{4}\rm\,K)^{-0.72} {\rm\,cm^{3}\,s^{-1}}$ in what
follows, although more accurate fits are available
\citep[e.g.][]{VernerFerland1996a}.  If the hydrogen is highly
ionized, then $n_{\rm HII}\simeq n_{\rm H}$, $n_{\rm e}\simeq n_{\rm
  H}(1+n_{\rm He}/n_{\rm H})$, and Eq.~(\ref{eq:eq}) may be rewritten
as
\begin{align} x_{\rm HI} &\simeq 9.6\times 10^{-6}\, \Delta \frac{(1+\chi_{\rm He})}{\Gamma_{-12}}\left(\frac{T}{10^{4}\rm~K}\right)^{-0.72}\left(\frac{\Omega_{\rm
      b}h^{2}}{0.022}\right) \nonumber \\ &\times
  \left(\frac{1-Y}{0.76}\right)\left(\frac{1+z}{5}\right)^{3}, \label{eq:xHI} \end{align}
where $x_{\rm HI}=n_{\rm HI}/n_{\rm H}$ is the \HI fraction,
$\Delta = \rho/\bar{\rho}$ is the fractional overdensity, 
$\chi_{\rm He}=\frac{\eta Y}{4(1-Y)}$ accounts for electrons released
by singly ($\eta=1$) and doubly ($\eta=2$) ionized helium\footnote{The
  reionization of neutral helium, with an ionization potential of
  $24.6\rm\,eV$, is thought to occur at the same time as \HI
  reionization \citep{Friedrich2012}.  In contrast, the reionization
  of singly ionized helium, with an ionization potential
  $54.4\rm\,eV$, is not expected to complete until $z\simeq2$--$3$ due
  to the relative scarcity of ionizing sources (most likely active
  galactic nuclei) with hard spectra at higher redshifts
  \citep{FurlanettoOh2008,McQuinn2009,Compostella2014}.} and
$\Gamma_{-12}=\Gamma_{\rm HI}/10^{-12}\rm\,s^{-1}$.

Finally, note that photo-ionization by the UVB also heats the low-density IGM \citep{MiraldaRees1994}.  Well after reionization,
photo-heating and cooling due to the adiabatic expansion of the
Universe produces a well defined temperature-density
relation\footnote{Immediately following (inhomogeneous) hydrogen
  reionization, the temperature-density relation is instead expected
  to exhibit considerable scatter
  \citep{Bolton2004,FurlanettoOh2009,LidzMalloy2014}, and may even be
  inverted, $\gamma<1$ \citep{Trac2008}.  Evidence for fluctuations in
  the thermal state of the IGM at high redshift may thus provide
  further, indirect evidence of the end stages of reionization.} for
gas overdensities $\Delta \lesssim 10$, where $T =
T_{0} \Delta^{\gamma -1}$ (\citealt{HuiGnedin1997,McQuinn2015}).
Typically, the temperature at mean density is $10^{3.7} \leq T_{0}
\leq 10^{4.3} \rm~K$, while the slope of the relation is $ 1 \leq
\gamma \leq 1.6$
\citep{Schaye2000,Ricotti2000,McDonald2001,Lidz2010,Becker2011,Garzilli2012,Rudie2012,Bolton2014,Boera2014}.
Combining this temperature-density relation with Eqs.~(\ref{eq:GP})
and (\ref{eq:eq}) gives
\begin{align} \tau_{\rm FGPA}^{\alpha} &\simeq 2.2 \,\Delta^{\beta} \frac{(1+\chi_{\rm
      He})}{\Gamma_{-12}}
  \left(\frac{T_{0}}{10^{4}\rm~K}\right)^{-0.72}\left(\frac{\Omega_{\rm
      b}h^{2}}{0.022}\right)^{2} \nonumber\\ &\times
  \left(\frac{\Omega_{\rm
      m}h^{2}}{0.141}\right)^{-1/2}\left(\frac{1-Y}{0.76}\right)^{2}\left(\frac{1+z}{5}\right)^{9/2},
\label{eq:FGPA}
\end{align}
where $\beta = 2 - 0.72(\gamma - 1)$.  This is the Fluctuating
Gunn-Peterson Approximation (FGPA, {\it e.g.}
\citealt{Rauch1997,Weinberg1999}).  It neglects redshift space
distortions and spatial fluctuations in the UVB, but elucidates the
dependence of the \Lya forest optical depth on underlying physical
quantities.  Importantly, this includes the photo-ionization rate
$\Gamma_{\rm HI}$ (and hence specific intensity $J(\nu,z)$) of the
  UVB.

\subsection{Measurements of the photo-ionization rate} \label{sec:PIrate}

With the relationship between the \Lya forest opacity and the UVB
intensity established, we now consider how measurements are made in
practice.  There are two main techniques employed in the literature to
measure the metagalactic photo-ionization rate, $\Gamma_{\rm HI}$:
modelling the mean transmitted flux in the \Lya forest
\citep{Rauch1997} and the quasar proximity effect
\citep{Murdoch1986,Bajtlik1988}.

The first approach typically relies on using numerical simulations of
structure formation to reproduce the observed mean transmission in the
\Lya forest, $\langle F \rangle = \langle e^{-\tau_{\alpha}} \rangle$.
Mock \Lya forest spectra are extracted from simulations performed
using a spatially uniform, time-dependent UVB model
\citep[e.g.][]{HaardtMadau1996,HaardtMadau2012,FaucherGiguere2009}.
The amplitude of the UVB model is then treated as an adjustable
parameter; since in photo-ionization equilbrium $\tau_{\alpha} \propto
\Gamma_{\rm HI}^{-1}$ (e.g. Eq.~\ref{eq:FGPA}), rescaling the
simulated \Lya optical depths is straightforward.  The
photo-ionization rate $\Gamma_{\rm HI}$ (and hence $J(\nu)$, if the
UVB spectral shape is known) may then be obtained by matching the mean
transmission in the simulated spectra to observational measurements at
$2<z<6$
\citep[e.g.][]{Kim2002,Schaye2003,Kirkman2005,Fan2006,FaucherGiguere2008b,Paris2011,Becker2013,Becker2015}.

This method has been widely applied to \Lya forest data at $z>2$
\citep{McDonaldMiraldaEscude2001,MeiksinWhite2003,Tytler2004,Bolton2005,FaucherGiguere2008,WyitheBolton2011,Rollinde2013,BeckerBolton2013}.
However, this approach will start to break down approaching $z=6$,
when the assumption of a spatially uniform UVB no longer holds (see
also \S\ref{sec:UVfluc}).  Corrections for the effect of spatial
fluctuations in the UVB are therefore applied to the measurements
\citep{MeiksinWhite2003,BoltonHaehnelt2007}.  The mean transmitted
flux furthermore depends on the IGM temperature and cosmology
(e.g. Eq.~\ref{eq:FGPA}). Independent constraints on these quantities
are therefore required, and the historical range of estimates for
$\Gamma_{\rm HI}$ in literature are in part attributable to differing
assumptions for these parameters.

The second widely used approach, the line-of-sight proximity effect,
estimates $\Gamma_{\rm HI}$ by quantifying the reduction in the
average \Lya opacity approaching the systemic redshift of a quasar
\citep{Weymann1981,Carswell1982}. The ionizing radiation from a quasar
will locally ionize the hydrogen in the IGM to a higher degree than
the UVB.  Ignoring peculiar velocites and spatial variations in gas
temperature, since $\tau_{\alpha}\propto \Gamma_{\rm HI}^{-1}$ it
follows that the \Lya optical depth as a function of proper distance
$r$ from a quasar is
\begin{equation} \tau_{\alpha}^{\rm UVB+Q} = \frac{\tau_{\alpha}^{\rm UVB}}{1+\Gamma_{\rm HI}^{\rm Q}(r)/\Gamma_{\rm HI}^{\rm UVB}}, \end{equation}
where $\tau_{\alpha}^{\rm UVB}$ is the typical optical depth in
the \Lya forest, $\Gamma_{\rm HI}^{\rm Q}(r)$ is the ionization rate of
the quasar and $\Gamma_{\rm HI}^{\rm UVB}$ is the UVB ionization rate.
If the absolute magnitude, redshift and spectral energy distribution
of the quasar are known, $\Gamma_{\rm HI}^{\rm Q}(r)$ may be computed with
Eq.~(\ref{eq:Gamma}).  Estimates for $\tau_{\alpha}^{\rm UVB+Q}$
and $\tau_{\alpha}^{\rm UVB}$ then lead to $\Gamma_{\rm HI}^{\rm
  UVB}$.

\begin{figure}
\begin{center}
\includegraphics[width=0.48\textwidth]{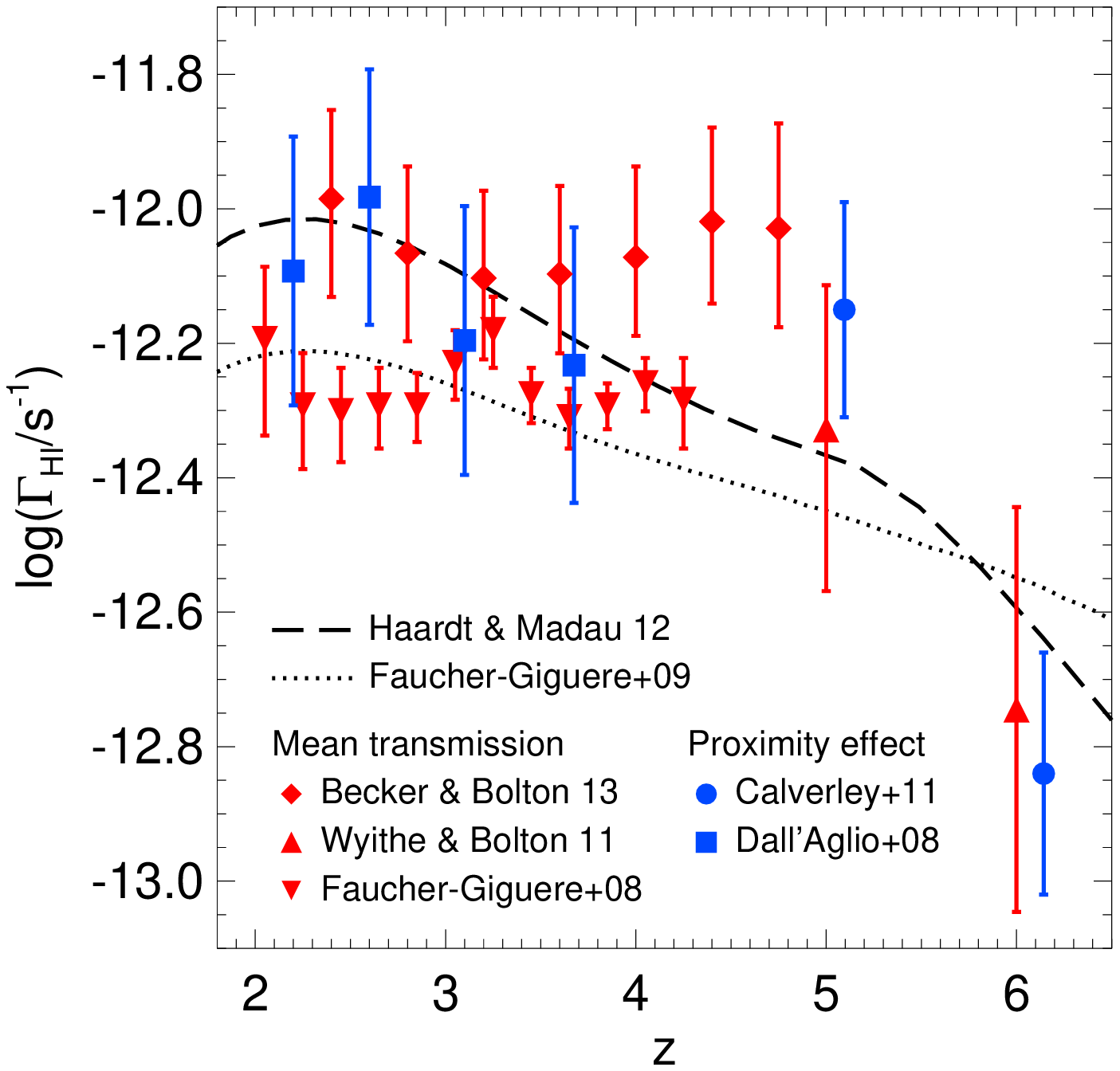}
\vspace{-1cm}
\caption{Summary of recent $\Gamma_{\rm HI}$ measurements obtained
  from the mean \Lya forest transmission
  \citep{FaucherGiguere2008,WyitheBolton2011,BeckerBolton2013} and the
  proximity effect \citep{Dallaglio2008,Calverley2011}.  Note that
  some of the data points have been offset by $\Delta z=0.05$ for
  clarity, and the \citet{Dallaglio2008} data have been rebinned and
  converted from a specific intensity assuming $\alpha_{\rm bg}=1.5$.
  These data are furthermore quoted directly from the literature.
    However, caution must be exercised with any direct comparison here
    as differing assumptions for systematic uncertainties in these
    studies such as e.g. the IGM temperature, can increase the scatter
    in the measurements.  For further discussion of this point see
    \citet{BeckerBolton2013}.  For comparison, the dashed and dotted curves
  display the empirically calibrated UVB models constructed by
  \citet{HaardtMadau2012} and \citet{FaucherGiguere2009},
  respectively.  These are based on the expected contribution to the
  UVB from star forming galaxies and quasars, and an empirical model
  for the ionizing opacity of the IGM. }
\label{fig:Gamma}
\end{center}
\end{figure}

The optical depth, however, is not a directly observable quantity;
early proximity effect analyses therefore focussed on measuring the
number density of \Lya absorption lines blueward of a quasar's \Lya
emission line
\citep[e.g.][]{Murdoch1986,Bajtlik1988,Giallongo1996,Scott2000}.  More
recent approaches have instead analysed the transmitted flux,
$F=e^{-\tau_{\alpha}}$, often combined with numerical
simulations of the IGM which model the density, temperature and
velocity field around quasar host haloes in detail
\citep{FaucherGiguere2008b,Dallaglio2008,Calverley2011}.  
In addition to modeling the atypical environment of quasars, care
must be taken to avoid further potential biases in the
measurements arising from line-of-sight variations in the IGM density
distribution 
\citep[e.g][]{LoebEisenstein1995,Rollinde2005,Guimaraes2007,Partl2011}.

The typical values obtained using both these techniques are
$\Gamma_{\rm HI}\sim 10^{-12}\rm\, s^{-1}$ at $2\leq z \leq 4$,
declining by approximately a factor of two and four approaching
reionization at $z=5$ and $z=6$, respectively.  Some recent
measurements are summarised in Fig.~\ref{fig:Gamma}.  Inferences about
the underlying ionizing source population can be made by comparing
these measurements to estimates of the emissivity from known ionizing
sources at $2<z<6$, obtained by integrating observed quasar and Lyman
break galaxy luminosity functions.  We discuss this comparison in more
detail in \S\ref{sec:gals}. However, this first requires converting
$\Gamma_{\rm HI}$ into an emissivity, which itself relies on estimates
for the typical mean free path for ionizing photons and its evolution
with redshift.

\subsection{The mean free path at the Lyman limit} \label{sec:mfp}

Consider the mean free path for ionizing photons in an IGM populated
by Poisson distributed \HI absorbers, with column densites $N_{\rm
  HI}$ described by the column density distribution function (CDDF),
$f(N_{\rm HI},z)= \partial^{2}{n}/\partial N_{\rm HI}\partial
z$. The intervening effective optical depth (see also Eq.~\ref{eq:Jnu}) for
photons with frequency $\nu_{0}$ at redshift $z_{0}$ that were emitted
at redshift $z$ is then \citep{Paresce1980}
\begin{equation} {\bar \tau}(\nu_{0},z_{0},z)=\int_{z_{0}}^{z}dz^{\prime}\int_{0}^{\infty}dN_{\rm HI}f(N_{\rm HI},z^{\prime})(1-e^{-\tau_{\nu}}), \label{eq:tau_ll} \end{equation}
where $\tau_{\nu}=\sigma_{\rm HI}(\nu)N_{\rm HI}$.  Parameterising the
CDDF as $f(N_{\rm HI},z) = N_{0}N_{\rm HI}^{-\beta_{\rm
    N}}(1+z)^{\beta_{\rm z}}$ and evaluating the integral yields
(e.g. \citealt{FaucherGiguere2008})
\begin{align} {\bar \tau} &=N_{0} \frac{\Gamma(2-\beta_{\rm N}) (1+z_{0})^{3(\beta_{\rm N}-1)}}{(\beta_{\rm N}-1)(\beta_{\rm z}-3\beta_{\rm N} + 4)}\left(\frac{\nu_{0}}{\nu_{\rm LL}}\right)^{-3(\beta_{\rm N}-1)}  \nonumber\\ &\times  \sigma_{\rm LL}^{\beta_{\rm N}-1}  \left[(1+z)^{\beta_{\rm z}-3\beta_{\rm N}+4}-(1+z_{0})^{\beta_{\rm z}-3\beta_{\rm N}+4}\right], \nonumber \end{align}
\noindent
where $\Gamma$ is the Gamma function.  The mean free path is then the
distance a photon can travel before encountering an optical depth of
unity.  Noting that $dl/d{\bar \tau}=(dl/dz)/(d{\bar \tau}/dz)\simeq
\lambda_{\rm mfp}$ when $\Delta{\bar \tau}=1$ thus leads to
\begin{align} \lambda_{\rm mfp}(\nu) &\simeq \frac{c(\beta_{\rm N}-1)}{N_{0}\sigma_{\rm LL}^{\beta_{\rm N}-1}\Gamma(2-\beta_{\rm N})}\left(\frac{\nu}{\nu_{\rm LL}}\right)^{3(\beta_{\rm N}-1)} \nonumber \\
 &\times \frac{1}{H_{0}\Omega_{\rm m}^{1/2}(1+z)^{\beta_{\rm
        z}+5/2}}, \label{eq:mfp} \end{align} which gives an analytical
approximation for the mean free path for ionizing photons at $z
\gtrsim 2$.  This may be written more compactly as $\lambda_{\rm
  mfp}=\lambda_{\rm LL}(z)(\nu/\nu_{\rm LL})^{3(\beta_{\rm N}-1)}$,
where $\lambda_{\rm LL}(z)$ is the mean free path at the Lyman limit.
In general, therefore, higher frequency photons have a larger mean
free path and $\lambda_{\rm LL}(z)$ decreases with increasing
redshift, but the precise normalisation, frequency and redshift
dependence of Eq.~(\ref{eq:mfp}) relies on an accurate observational
determination of the CDDF or a related quantity.

Many surveys have attempted to infer $\lambda_{\rm LL}(z)$ by
measuring $f(N_{\rm HI},z)$ and using Eq.~(\ref{eq:mfp}).  Critically,
however, the absorption systems which dominate the opacity, the
so-called Lyman limit systems (LLSs, $10^{17.2}\rm\,cm^{-2}\leq N_{\rm
  HI}\leq 10^{19}\rm\,cm^{-2}$) and saturated \Lya forest absorbers ($
10^{14.5} \rm\,cm^{-2}\leq N_{\rm HI}\leq 10^{17.2}\rm\,cm^{-2}$), are
those for which $N_{\rm HI}$ is most difficult to determine
observationally.  Extrapolations over this $N_{\rm HI}$ range are
therefore often employed.  Previous studies have generally found
$\beta_{\rm z}\simeq 1$--$3$ (note this is dependent on $N_{\rm HI}$)
and $\beta_{\rm N}\simeq 1.3$--$1.7$ at $z>2$
\citep{Tytler1987,Petitjean1993,StorrieLombardi1994,Kim1997,SongailaCowie2010,Rudie2013}.

However, recent work has demonstrated that the CDDF is not well
described by a single power-law index $\beta_{\rm N}$, and may be
better represented by a series of broken power-laws
\citep{Prochaska2010,Kim2013,OMeara2013}.  The expected clustering of
LLS will also impact on $\lambda_{\rm mfp}$ estimated from
Eq.~(\ref{eq:mfp}), as this expression assumes the absorbers are
Poisson distributed \citep{Prochaska2014}.  This has led to the
development of alternative approaches based on directly measuring the
optical depth at the Lyman limit, $\tau_{\rm LL}$, using stacked
spectra \citep{Prochaska2009,OMeara2013,fumagalli2013,Worseck2014}.
Another approach used for estimating $\lambda_{\rm mfp}$ is to use
hydrodynamical simulations which explicitly model the distribution of
gas in the IGM, coupled with radiative transfer or analytical
treatments for self-shielding
\citep{MiraldaEscude2000,MeiksinWhite2004,GnedinFan2006,BoltonHaehnelt2007,McQuinn2011,Emberson2013,SobacchiMesinger2014,Munoz2014}.

\begin{figure}
\begin{center}
\includegraphics[width=0.45\textwidth]{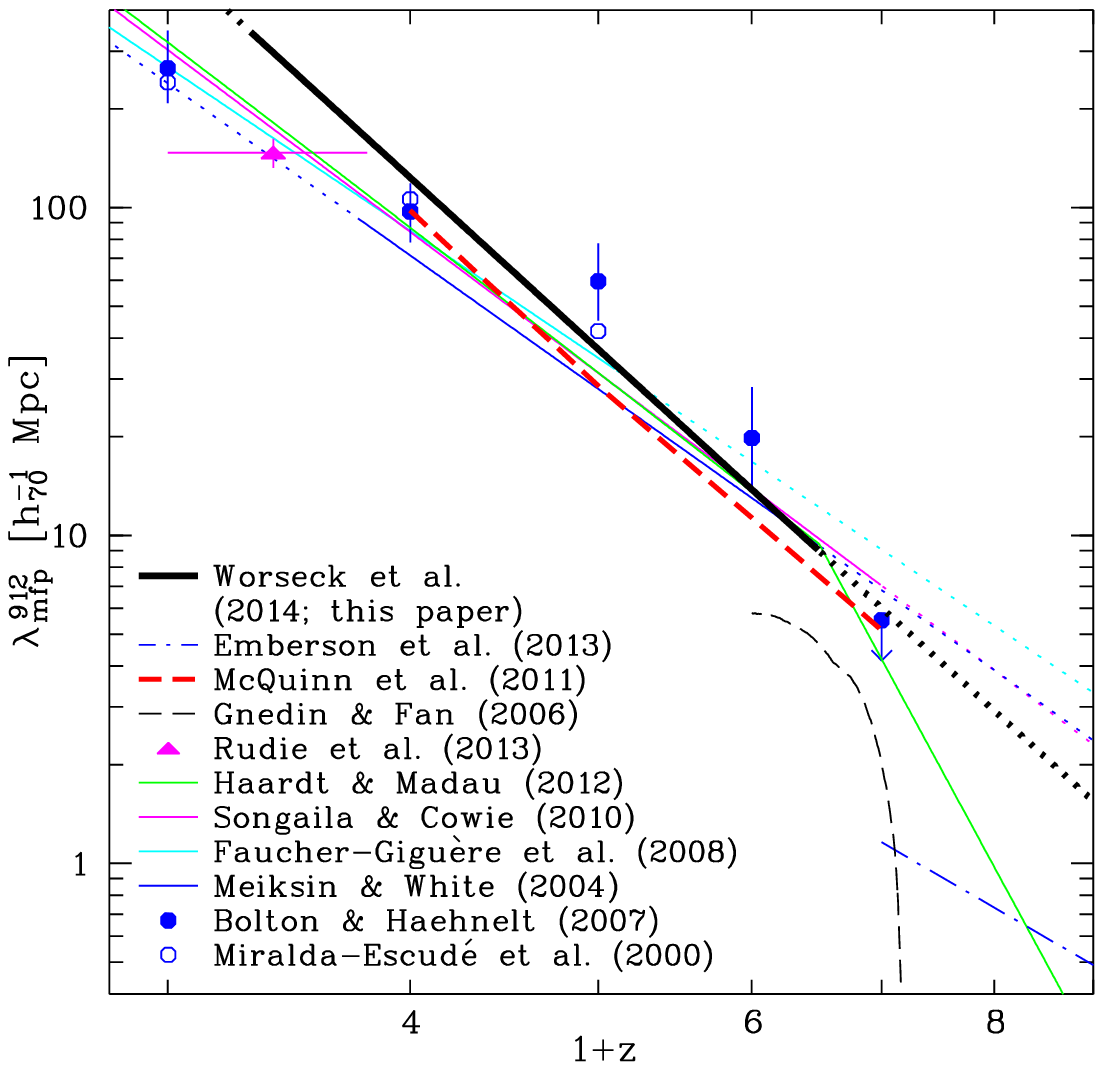}
\caption{Summary of estimates for the proper mean free path at the
  Lyman limit, $\lambda_{\rm LL}(z)$.  The results are based on
  stacked quasar spectra \citep{Worseck2014}, observations of discrete
  \HI absorption systems
  \citep{FaucherGiguere2008,SongailaCowie2010,HaardtMadau2012,Rudie2013},
  optically thin IGM simulations using either semi-analytical
  \citep{MiraldaEscude2000,BoltonHaehnelt2007} or empirical
  \citep{MeiksinWhite2004} corrections for the abundance of
  self-shielded absorbers, and full radiative transfer models
  \citep{GnedinFan2006,McQuinn2011,HaardtMadau2012,Emberson2013}.
  Note that the estimates at $z\gtrsim 6$ are either extrapolations
  from data at $z<6$ or predictions from simulations. {\it Reproduced
    from Figure 11 of \citet{Worseck2014} by permission of the authors.}}
\label{fig:mfp}
\end{center}
\end{figure}

A selection of recent measurements and simulation predictions are
shown in Fig.~\ref{fig:mfp}, from \cite{Worseck2014}.  A recent
analysis of $\tau_{\rm LL}$ in stacked quasar spectra by
\cite{Worseck2014} provides a best fit evolution of $\lambda_{\rm
  LL}=37[(1+z)/5]^{-5.4}$ proper Mpc at $2<z<5$.  Using an alternative
approach based on the observed incidence of LLSs,
\cite{SongailaCowie2010} instead find a slightly shallower evolution
with redshift, $\lambda_{\rm LL}=32[(1+z)/5]^{-4.4}$ proper Mpc.
Regardless of the precise normalisation and slope, however, these
results indicate $\lambda_{\rm mfp}$ evolves more quickly than
expected for a population of absorbers with no intrinsic
evolution.\footnote{The expected evolution in absorption line number
  density with redshift is $dn/dz = cn_{\rm
    abs}(z)\sigma_{\rm abs}(z)(1+z)^{2}/H(z)$, where $n_{\rm abs}(z)$
  is the comoving number density of absorbers and $\sigma_{\rm
    abs}(z)$ is the physical absorber cross-section.  At $z \geq 2$,
  $H(z)\propto (1+z)^{3/2}$ and thus $dn/dz \propto
  (1+z)^{1/2}$ if $n_{\rm abs}(z)$ and $\sigma_{\rm abs}(z)$ are held
  fixed (i.e. $\beta_{\rm z}=0.5$).  Hence from Eq.~(\ref{eq:mfp}),
  for a non-evolving population of Poisson distributed absorbers, the
  (proper) mean free path $\lambda_{\rm mfp}\propto (1+z)^{-3}$.}

Consequently, there is significant evolution in the comoving number
density and typical cross-section of Lyman limit absorbers toward
higher redshift.  This is directly related to the decline in the
photo-ionization rate at $z>5$; as the overall level of ionization in
the IGM drops, the typical density of LLS decreases, extending these
optically thick systems out from the central regions of dark matter
haloes to the surrounding filaments in the IGM.  It has been suggested
this coupling between $\lambda_{\rm mfp}$ and $\Gamma_{\rm HI}$ (or
equivalently, the sinks and sources of ionizing photons in the IGM)
naturally explains the flat evolution in $\Gamma_{\rm HI}$ from
$2<z<5$ and its sudden decline approaching $z=6$
\citep{McQuinn2011,Munoz2014} when in contrast, the comoving ionizing
emissivity evolves slowly.

\subsection{The ionizing emissivity}

Armed with observational determinations of $\Gamma_{\rm HI}(z)$ and
$\lambda_{\rm LL}(z)$ at $z<6$, constraints on the integrated
emissivity from ionizing sources may now be obtained. If the typical
mean free path for Lyman limit photons is much smaller than the
horizon scale, $\lambda_{\rm mfp}(\nu) \ll c/H(z)$, photon redshifting
effects are minimal and Eq.~(\ref{eq:Jnu}) may be written as
\begin{equation} J(\nu) \simeq \frac{\epsilon(\nu)\lambda_{\rm mfp}(\nu)}{4\pi} \label{eq:lsa}. \end{equation}  
This is referred to as the local source approximation
\citep{Madau1999,SchirberBullock2003}.  The proper specific
emissivity, $\epsilon(\nu)$, is related to the comoving emissivity of
ionizing photons by
\begin{equation} {\dot N}_{\rm ion}(z)=\frac{1}{(1+z)^{3}}\int_{\nu_{\rm LL}}^{\infty}d\nu\frac{\epsilon(\nu)}{h_{\rm P}\nu}\simeq \frac{\epsilon_{\rm LL}}{h_{\rm P}\alpha_{\rm s}(1+z)^{3}}, \label{eq:NioneLL} \end{equation}
where the simplifying assumption of a power-law source spectrum,
$\epsilon(\nu)=\epsilon_{\rm LL}(\nu/\nu_{\rm LL})^{-\alpha_{\rm s}}$
is adopted.  Substituting Eqs.~(\ref{eq:Gamma}), (\ref{eq:lsa}) and
(\ref{eq:mfp}) into this expression gives
\begin{equation} {\dot N}_{\rm ion}(z) \simeq \frac{\Gamma_{\rm HI}(z)}{\sigma_{\rm LL}\lambda_{\rm LL}(z)} \frac{(\alpha_{\rm bg}+3)}{\alpha_{\rm s}} \frac{1}{(1+z)^{3}}, \label{eq:Nion} \end{equation}
where we may also identify $\alpha_{\rm bg}=\alpha_{\rm
  s}-3(\beta_{\rm N}-1)$ if the CDDF is described by a single power
law index (but see \S\ref{sec:mfp}).

Eq.~(\ref{eq:Nion}) may be used to estimate the total ionizing
emissivity in the IGM, given observational determinations of
$\Gamma_{\rm HI}(z)$ and $\lambda_{\rm LL}(z)$.  Values for
$\alpha_{\rm s}$ will depend on the sources which dominate the UVB.
For active galactic nuclei (AGN), $\alpha_{\rm s}=1.5$--$1.8$
\citep{Telfer2002,Stevans2014}, whereas for star forming galaxies
$\alpha_{\rm s}=1$--$3$
\citep[e.g.][]{Leitherer1999,EldridgeStanway2012}.  Note, however, the
local source approximation ignores the cosmological redshifting of
ionizing photons to frequencies $\nu<\nu_{\rm LL}$, and thus
underestimates $\dot{N}_{\rm ion}$ for a given $\Gamma_{\rm HI}$ by
$\sim 50\%$ ($\sim 10\%$) at $z=2$ ($z=5$) \citep{BeckerBolton2013}.
Modelling $\lambda_{\rm LL}$ with Eq.~(\ref{eq:mfp}) and adopting a
single power-law source spectrum are further simplifications; these
enable analyical forms for the integrals, but more detailed treatments
often require these equations to be solved numerically.

Recent inferences are consistent with ${\dot N_{\rm ion}}\sim
10^{51}\rm\,s^{-1}\,Mpc^{-3}$ at $2<z<6$ \citep{BeckerBolton2013}.
These are around a factor of two higher than earlier estimates
\citep{BoltonHaehnelt2007,KuhlenFaucherGiguere2012}, largely due to
improved constraints for the mean free path at $z>4$ and the
temperature of the IGM at $2<z<5$.  The emissivity is also often
expressed as the number of ionizing photons emitted per hydrogen atom
over a Hubble time, ${\dot n}_{\rm ion}={\dot N}_{\rm
  ion}(1+z)^{3}/{\bar n}_{\rm H}H(z)$, where $t_{\rm age}\simeq
2/3H(z)$ is the age of the Universe at $z \geq 2$.
\citet{MiraldaEscude2003} first pointed out that the emissivity at
$z=4$ corresponds to $\dot{n}_{\rm ion}\leq 7$, which strictly limits
the amount by which it may decline at $z>4$ if reionization is to
complete by $z=6$.  Similar conclusions were reached by
\citet{Meiksin2005}, and later \citet{BoltonHaehnelt2007}, who
extended the analysis to $z=6$, finding ${\dot n}_{\rm ion} \leq
2$--$5$.

In comparison, recent cosmological radiative transfer simulations
indicate the reionization of the IGM requires $\gtrsim$2-3 ionizing
photons per atom to counterbalance radiative recombinations
\citep{Finlator2012,So2014}. The close correspondence between this
number and ${\dot n}_{\rm ion}$ indicates reionization may be a
``photon-starved'' process; there are only just enough photons present
to reionize the IGM at $z\simeq 6$.  Consequently, the metagalactic
ionizing emissivity at $z \leq 6$ is an important observational
constraint which viable reionization models at $z>6$ must anchor to
\citep{Pritchard2010,Alvarez2012,Finlator2012,Mitra2013,Fontanot2014}.

\subsection{The contribution from galaxies and AGN} \label{sec:gals}

\begin{figure*}
   \centering
   \begin{minipage}{\textwidth}
   \begin{center}
      \includegraphics[width=0.9\textwidth]{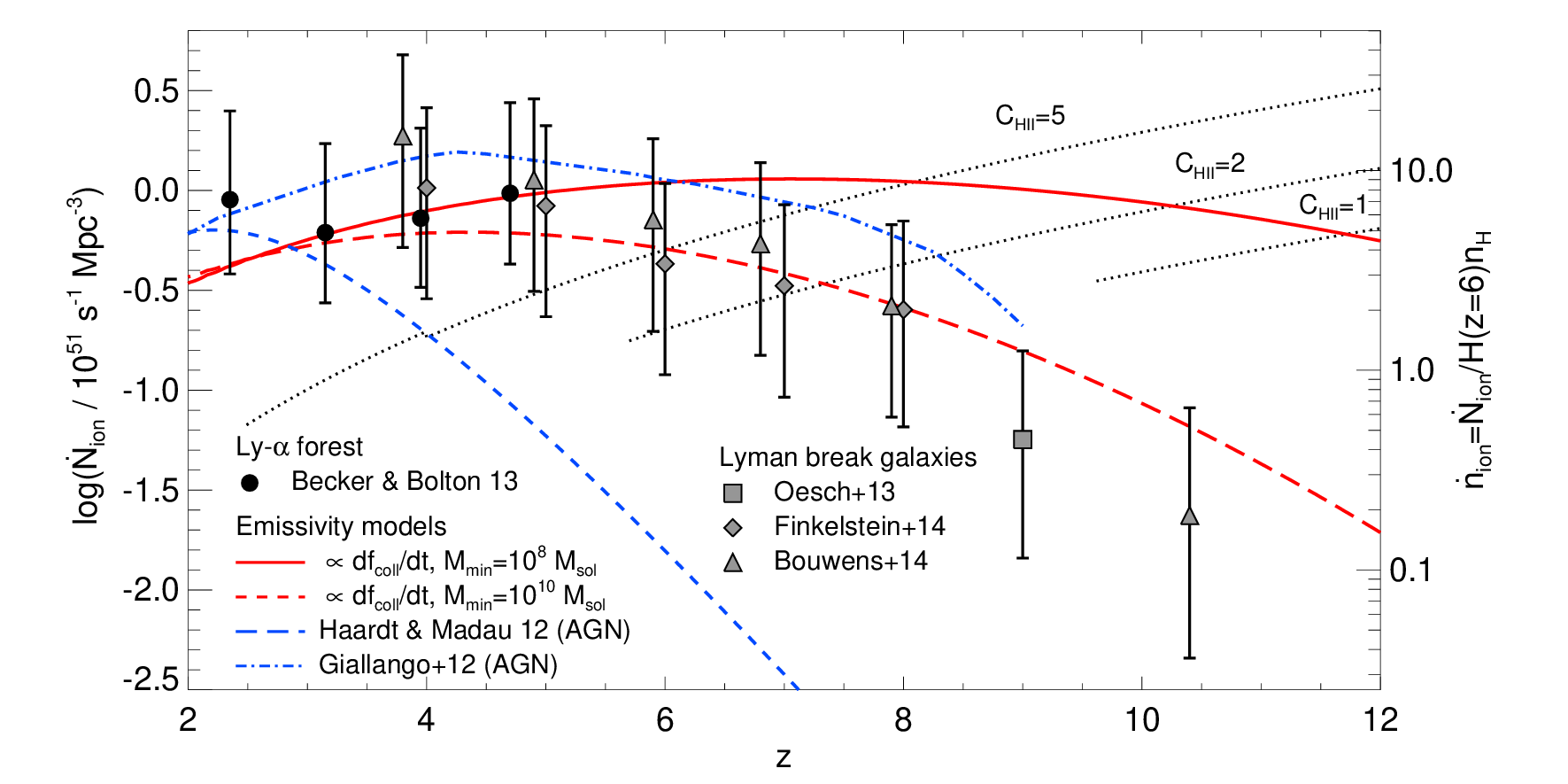}
      \vspace{-0.4cm}
      \caption{The comoving emissivity of ionizing photons with
        redshift.  The filled circles correspond to estimates from the
        \Lya forest opacity \citep{BeckerBolton2013} (see their
          table 3 for a full breakdown of the systematic
          uncertainties), and the grey symbols show results
          derived here from the comoving UV luminosity density,
        $\rho_{\rm UV}$, of Lyman break galaxies
        \citep{Oesch2013,Finkelstein2014,Bouwens2014}.  The error
          bars on these data are computed by adding the published
          uncertainies on $\rho_{\rm UV}$ in quadrature with the
          systematic uncertainies $f_{\rm LL}=0.25\pm0.10$, $f_{\rm
            esc}=0.3\pm0.2$ and $\alpha_{\rm s}=2\pm1$ (see text and
          Eqs~(\ref{eq:NioneLL}) and~(\ref{eq:eUV}) for details). The red curves
        display a model for the emissivity from star forming galaxies,
        where the ionizing efficiency is fixed at $\zeta=25$
        independent of redshift.  Here ${\dot N}_{\rm ion}$ is
        proportional to the rate of change of the collapsed fraction
        of matter in haloes with $M_{\rm min}\geq 10^{8}M_{\odot}$
        (solid curve) and $M_{\rm min}\geq 10^{10}M_{\odot}$ (long
        dashed curve).  The blue short-dashed and dot-dashed curves
        show the AGN emissivity models from \citet{HaardtMadau2012}
        and \citet{Giallongo2012}, respectively.  The dotted curves
        correspond to the ionizing emissivity required to balance
        recombinations for three different hydrogen clumping factors,
        $C_{\rm HII}=1$, $2$ and $5$; these are truncated where the
        recombination timescale exceeds the age of the Universe.}   \label{fig:Nion}
   \end{center}
   \end{minipage}
\end{figure*}

The ionizing emissivity inferred from the \Lya forest may be directly
compared to estimates based on the observed population of sources at
high redshift.  The power of this approach is that the metagalactic
emissivity represents the combined photon output from {\em all
  ionizing sources}, irrespective of uncertain physics at galactic
scales.  If the UV luminosity function, $\phi(L_{\rm UV},z)$, for a
given source population is known, the proper UV specific
emissivity\footnote{This quantity is typically described in the
  observational literature as a comoving UV luminosity density,
  $\rho_{\rm UV}=\epsilon_{\rm UV}/(1+z)^{3}$ with units $\rm erg\,
  s^{-1} \,Hz^{-1} \,Mpc^{-3}$.  This is often subsequently converted
  to a star formation rate density (in units of $\rm M_\odot\, yr^{-1}\,Mpc^{-3}$), using the relation $\rho_{\rm
    SFR}=1.25\times10^{-28}\rho_{\rm UV}$
  \citep{Kennicutt1998,Finkelstein2014}.  This assumes a Salpeter
  initial mass function and a constant star formation rate.} is given
by
\begin{equation}\epsilon_{\rm UV}=(1+z)^{3}\int_{L_{\rm min}}^{\infty}dL_{\rm UV}L_{\rm UV}\phi(L_{\rm UV},z). \label{eq:eUV}\end{equation}
The luminosity function for Lyman-break galaxies is typically
parameterised as a \citet{Schechter1976} function, whereas AGN are
often described by a double power-law
\citep[e.g.][]{Ueda2003,Richards2006,Croom2009}. The proper emissivity
at the Lyman-limit is then $\epsilon_{\rm LL}=f_{\rm esc}f_{\rm
  LL}\epsilon_{\rm UV}$, where $f_{\rm LL}=\epsilon_{\rm
    LL,int}/\epsilon_{\rm UV}$ is the ratio of the intrinsic
  Lyman limit and UV emissivities, and $f_{\rm esc}$ is the (average)
fraction of ionizing photons which escape the local interstellar
medium.  For AGN, a broken power-law approximation is often used to
estimate $f_{\rm LL}$ and $f_{\rm esc}=1$ is typically assumed
\citep[e.g.][]{Madau1999}, while for high-redshift galaxies stellar
population synthesis models are employed to obtain $f_{\rm LL}$
and $f_{\rm esc}$ is a free parameter.  The spectral properties of
galaxies may also be constrained further with observed UV spectral
slopes
\citep{Finkelstein2012,Dunlop2013,Robertson2013,Bouwens2014b,DuncanConselice2015}.
The comoving emissivity of ionizing photons is then obtained using
Eq.~(\ref{eq:NioneLL}).

Fig.~\ref{fig:Nion} displays ${\dot N}_{\rm ion}$ inferred from the UV
luminosity densities reported by \citet{Oesch2013},
\citet{Bouwens2014} and \citet{Finkelstein2014},
assuming\footnote{Note the ${\dot N}_{\rm ion}$ measurements
      from the \Lya forest are independent of these (uncertain)
      assumptions and are thus complementary to constraints from the
      UV luminosity density.}  $f_{\rm LL}=0.25\pm0.10$, $f_{\rm
    esc}=0.3\pm 0.2$ and a power law spectral index below the
  Lyman-limit, $\alpha_{\rm s}=2\pm 1$
  \citep[e.g][]{BeckerBolton2013}.  These data are based on recent
measurements of the Lyman break galaxy luminosity function to a
limiting absolute magnitude of $M_{\rm UV}\simeq -17$ at $4<z<10.5$,
and are compared to independent measurements of ${\dot N}_{\rm ion}$
from the \Lya forest at $2<z<5$, along with models for the expected
emissivity from AGN (blue curves) and star forming galaxies (red
curves).  The dotted curves display the comoving emissivity required
to balance radiative recombinations in the IGM \citep{Madau1999},
where
\begin{equation} {\dot N}_{\rm rec}= {\bar n}_{\rm H}^{2}(1+\chi_{\rm He})C_{\rm HII}(z)\alpha_{\rm HII}(T)(1+z)^{3}.  \label{eq:Nrec} \end{equation}
Here $C_{\rm HII}=\langle n_{\rm HII}^{2} \rangle/{\bar n}_{\rm
  HII}^{2}$ is the clumping factor.  This takes into account the
enhancement in the volume averaged recombination rate due to small
scale, dense structures.  Typical values for the ionized, low-density
IGM obtained from numerical simulations are $C_{\rm HII}\simeq 1$--$5$
\citep{Pawlik2009,Finlator2012,Shull2012,JeesonDaniel2014,So2014,KaurovGnedin2014}.
Alternatively, assuming the IGM is in ionization equilibrium
(cf. Eq.~\ref{eq:eq}), then Eq.~(\ref{eq:Nion}) and (\ref{eq:Nrec})
may be equated and rearranged for $C_{\rm HII}$.  Inserting
observational estimates for $\Gamma_{\rm HI}$ and $\lambda_{\rm LL}$
yields $C_{\rm HII}\simeq 2$--$3$
\citep{BoltonHaehnelt2007,McQuinn2011}.

Fig.~\ref{fig:Nion} demonstrates that the observed population of Lyman
break galaxies at $z<6$ is consistent with the emissivity inferred
from the \Lya forest, although the lack of evolution in the latter
from $z=4$--$5$ implies that the escape fraction of ionizing photons,
$f_{\rm esc}$, may be increasing toward higher redshift
\citep[e.g.][]{Alvarez2012,Ciardi2012,KuhlenFaucherGiguere2012,Shull2012,FerraraLoeb2013}.
However, the emissivity from these galaxies rapidly drops below the
critical rate required to balance recombinations at $z\gtrsim 6$,
indicating the observed sources are insufficient to drive reionization
to completion.  There are three ways to resolve this dilemma: (i)
there are many more faint galaxies below the detection limit which
contribute to the total photon budget; (ii) the ionizing efficiency
(the product of $f_{\rm LL}$ and $f_{\rm esc}$) of these galaxies
increases significantly toward higher redshift or (iii) there is
another population of sources which produce ionizing photons.  A
combination of all three may also be plausible.

To illustrate this further, the red curves in Fig.~\ref{fig:Nion}
display a simple model for the expected emissivity from star forming
galaxies, where
\begin{equation} \dot{N}_{\rm ion}(z) =\zeta(z){\bar n}_{\rm b}\frac{df_{\rm coll}(>M_{\rm min},z)}{dt}, \end{equation}
\citep[e.g][]{Pritchard2010,Wyithe2010,SobacchiMesinger2014}.  Here
$f_{\rm coll}$ is the collapsed fraction of matter in haloes above a
minimum mass threshold $M_{\rm min}$ \citep{ShethTormen2002}, and
$\zeta(z)=f_{\star}f_{\rm esc}N_{\gamma}$ is the ionizing efficiency
of star forming galaxies (assumed to be constant, $\zeta=25$, in
Fig.~\ref{fig:Nion}).  This ionizing efficiency is itself a function
of three uncertain parameters, the star formation efficiency
$f_{\star}$, the number of ionizing photons produced per baryon within
stars, $N_{\gamma}$, and the escape fraction.\footnote{Illustrative
  values are $N_{\gamma}=4600$ ($17500$) for a Salpeter (top-heavy)
  IMF \citep{Wyithe2010}, $f_{\star}=0.01$--$0.1$ \citep{Behroozi2013}
  and $f_{\rm esc}=0.05$--$0.5$ \citep{Wise2014}, where the latter two
  values will vary with halo mass.}

Considering all haloes above the mass threshold at which atomic
cooling ceases to be efficient, $M_{\rm min}\simeq 10^{8}M_{\odot}$,
produces an emissivity in excess of that observed from galaxies with
$M_{\rm UV} \lesssim -17$ at $z\gtrsim 6.5$.  Better
  agreement may be obtained by extrapolating the observed faint end
of the luminosity function to $M_{\rm UV}\simeq -10$ with
Eq.~(\ref{eq:eUV}); faint, high-redshift dwarf galaxies are therefore
widely postulated as likely candidates for making up the observed
ionizing photon shortfall at $z>6$
\citep{Trenti2010,Robertson2013,Robertson2015,Fontanot2014,Bouwens2015}.
This illustrates the important role of IGM studies in providing a
complete census of early galaxy populations; many of the ionizing
sources are too faint to detect individually, but we can nevertheless
observe their combined impact on the surrounding intergalactic gas.
Alternatively, as discussed above the ionizing efficiency may increase
with redshift, if e.g. $f_{\rm esc}$ evolves with redshift.  However,
direct measurements of $f_{\rm esc}$ approaching reionization are
prohibited by the high opacity of the intervening IGM, and searches
for analogues at lower redshift indicate the amount of Lyman continuum
radiation escaping from galaxies is modest
\citep[e.g.][]{Vanzella2010,Boutsia2011,Nestor2013,Mostardi2013}.

Lastly, it remains possible that other sources contribute to the total
ionizing emissivity.  The most natural candidate are AGN, with hard,
non-thermal spectra which produce an abundance of ionizing photons.
The blue curves in Fig.~\ref{fig:Nion} display two very different
models for the predicted contribution of AGN to $\dot{N}_{\rm ion}$.
The lower estimate is from the UVB model of \cite{HaardtMadau2012},
which is based on the bolometric luminosity functions presented by
\cite{Hopkins2007} at $0<z<6$.  There is a sharp drop in ${\dot
  N}_{\rm ion}$ at $z>4$ as the number density of sources falls,
implying that AGN are unlikely to be the dominant sources powering
reionization \citep[see also][]{Cowie2009}.  On the other hand,
semi-analytical modelling by \cite{Giallongo2012}, incorporating faint
($M_{\rm UV}\simeq -18.5$) AGN, produces $\dot{N}_{\rm ion}$ estimates
up to two orders of magnitude larger.  If such faint AGN are
widespread, these could provide almost all the ionizing photons
required to complete reionization.  However, uncertainties associated
with the number of faint AGN at $z>4$ remain considerable, leaving
their precise contribution to the ionizing photon budget a matter of
debate \citep[e.g.][]{Glikman2011,Masters2012,Giallongo2015}.  The
number and relative contribution of AGN to the UVB must furthermore
remain consistent with upper limits on the unresolved X-ray background
\citep{Dijkstra2004,HaardtSalvaterra2015} and the spectral shape of
the UVB inferred from metal ion absorption lines (\S\ref{sec:metals})
at $2<z<4$ \citep{Agafonova2007,Fechner2011,boksenberg2015}.  Lastly,
the double reionization of helium is expected to compete around $z
\simeq 2$--$3$, based on observations of the intergalactic \HeII
opacity \citep{Shull2010,Worseck2011} and the IGM temperature
\citep{Schaye2000,Lidz2010,Becker2011}.  This is thought to be driven
by the hard photons emitted by AGN
\citep{FurlanettoOh2008,McQuinn2009}.  Consequently, their
contribution at $z>4$ must avoid a premature end to \HeII reionization
\citep{McQuinn2012}.

\subsection{Spatial fluctuations in the UVB} \label{sec:UVfluc}

Finally in this section, we turn to consider spatial inhomogeneities
in the UVB.  In addition to the average photo-ionization rate
determined from the mean transmission of the \Lya forest, one can also
examine spatial fluctations in $\Gamma_{\rm HI}$ using the
line-of-sight variation in the transmission as a function of redshift
\citep{Fan2006}.  If reionization is incomplete this may boost the
amplitude of spatial variations in the transmission, with some
lines-of-sight crossing mostly through ionized bubbles, and others
intersecting multiple neutral regions in the IGM
\citep{WyitheLoeb2006}. Note, however, the scatter in the average
transmission may also be large shortly after reionization completes,
when the local mean free path to ionizing photons is similar to
the typical separation between (clustered) sources.  In this
scenario, the mean free path itself may vary spatially; even though
the IGM is highly ionized, the UVB will exhibit significant spatial
fluctuations
\citep[e.g.][]{Zuo1992,MeiksinWhite2004,MesingerFurlanetto2009}.

Importantly, these fluctuations potentially yield information on the
timing of patchy reionization and the distribution of the ionizing
sources; the scatter in the average transmission may be an interesting
diagnostic in this respect.  However, fluctuations owing to spatial
variations in the underlying density field may also contribute
significantly to the observed scatter \citep{Lidz2006}.  The aliasing
of transverse, small scale density fluctuations in three dimensions to
larger scales along the one-dimensional quasar sightlines can further
obscure effects from incomplete reionization
\citep[cf.][]{KaiserPeacock1991}.

Nevertheless, recent work indicates that the observed scatter is
indeed larger than expected in uniform UVB models, especially at $z
\gtrsim 5.5$ \citep{Becker2015}.  This finding is driven in part by
the discovery of a contiguous $110h^{-1}$ comoving Mpc dark region in
the $z \sim 5.7$ \Lya forest.  Such a large scatter in the mean
transmission is seen at $z<6$ in radiative transfer simulations where
reionization competes between $z=6$--$7$ \citep{Gnedin2014}, and may
be a natural consequence of spatial variations in the mean free path
at the tail-end of reionization \citep{FurlanettoOh2005}.  It has also
been recently suggested the fluctuations may instead be indicative of
photo-ionization by rare, bright quasars
\citep{Chardin2015}. Alternatively, reionization may be incomplete at
$z \sim 5.5-6$ (\citealt{Lidz2007, Mesinger2010}, and see \S
\ref{sec:meanf_a}).  Consequently, while these observations are highly
suggestive, we have yet to determine their full implications for the
reionization history of the Universe.

\section{METAL ABSORPTION LINES}
\label{sec:metals}

Metal absorption lines complement the \Lya forest as probes of the
high-redshift Universe in that they reflect the chemical enrichment
and ionization state in the regions in and around galaxies.  The
overall abundance of metals reflects the integrated global star
formation history, while the number densities, ionic ratios, and
kinematic profiles of absorbers contain information on the mechanisms
by which galaxies produce and expel metals (i.e., feedback
mechanisms), the stellar populations that produced the metals, and the
ionization state of the metal-enriched gas.  Metal absorbers are
particularly valuable tracers of galaxies during reionization, as they
can arise from faint galaxies that are below the detection thresholds
of direct galaxy surveys.  In this section we give an overview of
recent results on metal lines over $z \sim 5-6$.  We first describe
the observations, then examine what metal absorbers tell us about
galaxies and their environments at these redshifts.

\subsection{Observations of metal lines near $z \sim 6$}

\begin{figure*}
   \centering
   \begin{minipage}{\textwidth}
   \begin{center}
      \includegraphics[width=0.9\textwidth]{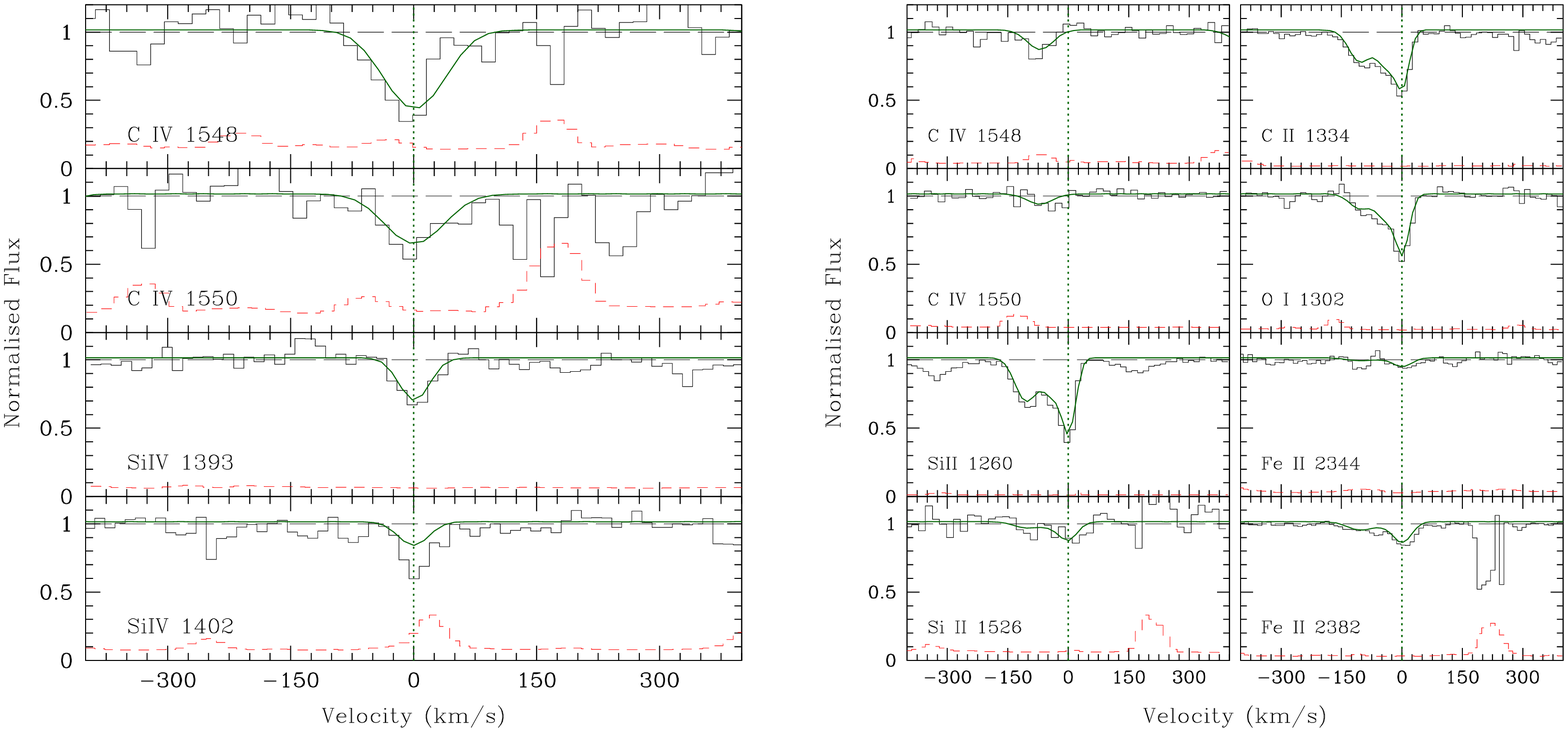}
      \vspace{-0.2cm}
      \caption{Examples of \civ\ (left) and low-ionization (right)
        metal absorption-line systems observed with VLT/X-Shooter, at
        $z=5.92$ and 5.79, respectively.  The transitions for a given
        absorber have been shifted onto a common velocity scale.
        Histograms show the continuum-normalized flux, while the
        dashed lines are the flux uncertainty.  Solid lines are Voigt
        profile fits.  {\it Reproduced from Figures 5 and 17 of \citet{dodorico2013} with
          permission of the authors.}}
   \label{fig:Nmetals}
   \end{center}
   \end{minipage}
\end{figure*}

Studies of metal lines near reionization have tended to focus on three
types of absorbers: highly ionized systems traced by \civ\ and/or
\siiv\ (which we will generally refer to as \civ\ systems);
low-ionization systems traced by \oi, \cii, \siii, and \feii\ (which
we will refer to as \oi\ systems); and \mgii\ systems, which can trace
a range of ionization states\footnote{Throughout this review we will
  follow the common convention in quasar absorption line literature of
  referring to an ion by its absorption spectrum, e.g., \civ\ for
  C$^{3+}$ ions.}.  Studies of \civ\ (rest-frame $\lambda =
1548,1551$~\AA) and \mgii\ ($\lambda = 2796,2804$~\AA), which are
observed in the infrared at $z > 5.5$ and 2.6, respectively, have
particularly benefited from the substantial increase in infrared
spectroscopic sensitivity recently provided by the X-Shooter
spectrograph on the Very Large Telescope \citep{sdodorico2006}, and
the Folded-port Infrared Echellette (FIRE) spectrograph on Magellan \citep{simcoe2013}.  We will begin
by describing the observations of \civ, \oi, and \mgii\ absorbers
separately, although the reader should bear in mind that these often
trace different components of the same system.  To date, quasars have
been used for systematic surveys of metals at $z > 5$; however, gamma
ray burst (GRB) afterglows have also started to yield samples of metal
lines at these redshifts
\citep[e.g.][]{Chornock2013,castrotirado2013,hartoog2014}.

\subsubsection{Metrics}

Before describing the observations we first introduce some of the
metrics used to quantify metal absorption line samples.  An absorber
population can be characterized by its column density distribution,
$f(N) = \partial^2 n/\partial N \partial X$, which gives the number of
systems per unit column density, $N$, and absorption path length
interval $X$ (see also \S\ref{sec:mfp}).  The absorption path length
interval is related to the redshift interval by
\begin{equation}
   \frac{dX}{dz} = (1+z)^2 \frac{H_{0}}{H(z)},
   \label{eq:X}
\end{equation}
\citep{bahcall1969}, and has the useful property that a population of
sources with a fixed physical cross-section will have a constant
number density per unit $X$.

Two quantities related to $f(N)$ that are often quoted for metal
absorbers are the line-of-sight number density, $dn/dX$ (or $dn/dz$),
and the comoving mass density.  The total mass density of an ion can
be expressed as a fraction of $\rho_{\rm crit}$, the critical density
at $z=0$,
\begin{equation}
   \label{eq:omega}
   \Omega_{\rm ion} = \frac{H_0 \,m_{\rm ion}}{c \, \rho_{\rm crit}} 
                    \int_{N_{\rm min}}^{N_{\rm max}} {N_{\rm ion} \, f(N_{\rm ion}) \, dN_{\rm ion}} \, ,
\end{equation}
which can be approximated as
\begin{equation}
   \label{eq:omega_approx}
   \Omega_{\rm ion} = \frac{H_0 \,m_{\rm ion}}{c \, \rho_{\rm crit}}
                    \frac{\sum{N_{\rm ion}}}{\Delta X} \, .
\end{equation}
Here, $N_{\rm ion}$ is the column density of an absorber and $\Delta
X$ is the total survey path length.

\subsubsection{\civ}

The \civ\ doublet is perhaps the most commonly used absorption-line
tracer of metals due to its sensitivity to enriched, highly ionized
gas.  The first \civ\ measurements at $z \sim 6$ were made by
\citet{simcoe2006b} and \cite{rw2006}, with larger samples and
improved sensitivity provided by \citet{becker2009}, \citet{rw2009},
\citet{simcoe2011a}, and \citet{dodorico2013}.  At present, 13 $z \sim
6$ quasars have been surveyed for \civ\ along their line of sight.  An
example system at $z=5.9$ is shown in Fig.~\ref{fig:Nmetals}, which
in this case displays both \civ\ and \siiv\ absorption.

While precise fits to $f(N_{\rm C\, IV})$ at $z \sim 6$ remain
difficult due to the small number of absorbers, \citet{dodorico2013}
find that the $z > 5.3$ data can be fit with a power law in $N_{\rm
  C\, IV}$ with a slope similar to that for \civ\ populations at lower
redshifts, but with a normalization that is a factor of 2 to 4 lower
\citep[see also][]{becker2009}.

\begin{figure}
   \begin{center}
   \includegraphics[width=0.45\textwidth]{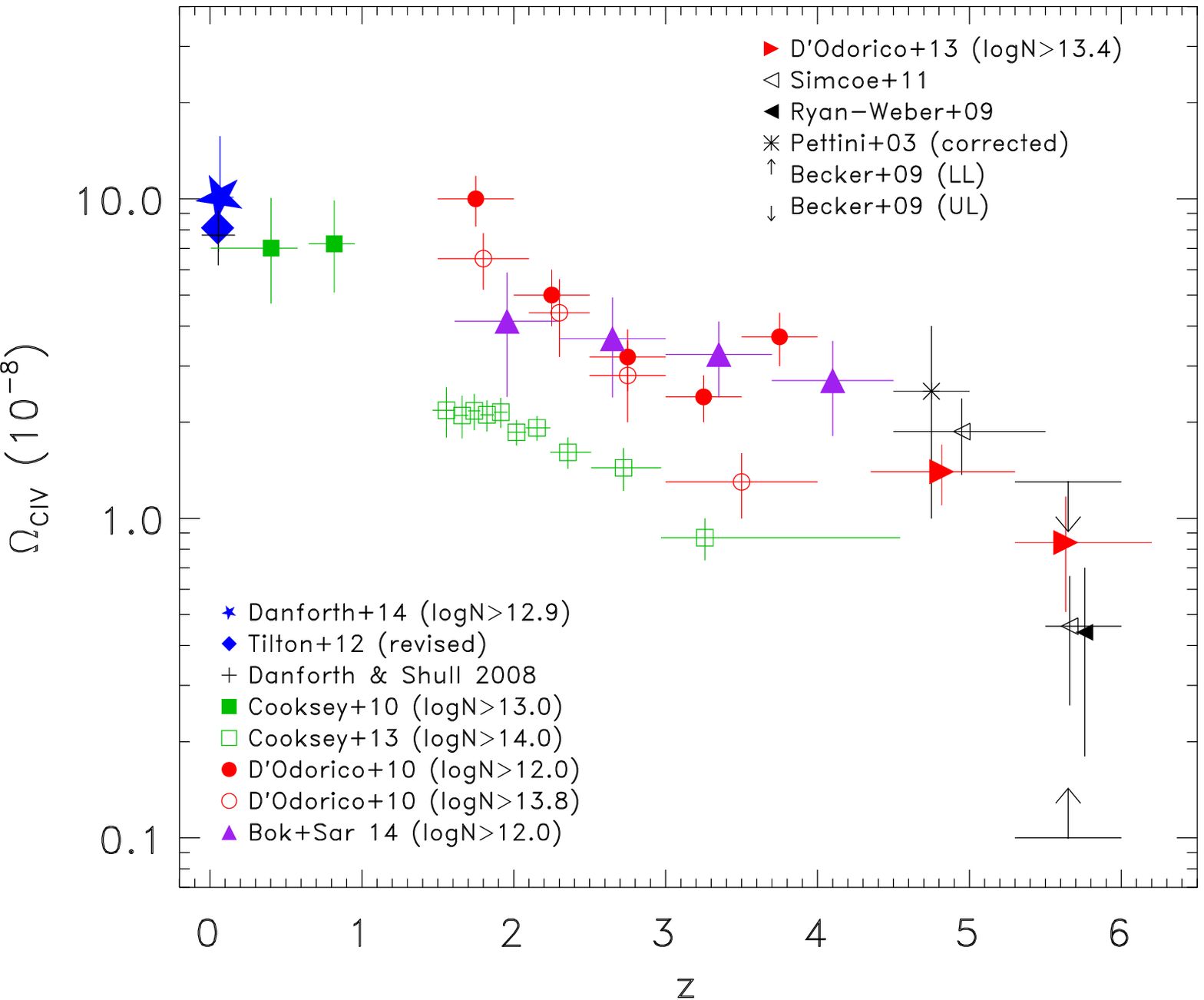}
      \vspace{-0.2cm}
      \caption{Evolution of the comoving mass density of \civ,
        expressed as a fraction of the critical density.  Results from
        the literature are plotted with symbols indicated in the
        figure legend \citep[][where ``Bok+Sar 14'' in the caption
          refers to Boksenberg \& Sargent
          2015]{pettini2003,danforth2008,becker2009,rw2009,cooksey2010,cooksey2013,
          dodorico2010,dodorico2013,simcoe2011a,tilton2012,danforth2014,boksenberg2015}.
        Note that some of the variation in these results is due to
        differences in the column density range over which
        $\Omega_{\rm C\,IV}$ is integrated.  {\it Reproduced from Figure 1 of
          \citet{shull2014} by permission of the authors and the AAS.}}
   \label{fig:Omega_CIV}
   \end{center}
\end{figure}

Fig.~\ref{fig:Omega_CIV} shows the evolution for $\Omega_{\rm C\, IV}$
over $0 < z < 6$.  In general, the mass density of \civ\ increases
towards lower redshifts, which is consistent with the buildup of
highly ionized metals in the circumgalactic medium of galaxies due to
processes such as outflows \citep[e.g.][]{oppenheimer2006}.  Changes in the ionization state of the metal-enriched gas will also affect the \civ\ statistics, however.  The
ionization state of \civ-selected gas will tend to evolve due to
changes in density and/or the UVB, such that the fraction of carbon in
these systems traced by \civ\ will change with redshift.  The
evolution of $\Omega_{\rm C\, IV}$ is therefore the product of
increasing metal enrichment and changes in the ionization state of the
metal-enriched gas.  Generally speaking, \civ\ becomes a preferred
ionization state at lower overdensities towards higher redshifts
\citep[e.g.][]{oppenheimer2006,oppenheimer2009}.  It is noteworthy
that $\Omega_{\rm C\, IV}$ appears to increase by a factor of two to
four from $z \sim 6$ to 5.  This suggests a substantial buildup of
metal-enriched regions around galaxies over this interval, although
the rate of increase may depend partly on the range of column
densities being considered \citep{dodorico2013}.

\subsubsection{\oi}

\begin{figure}
   \begin{center}
   \includegraphics[width=0.45\textwidth]{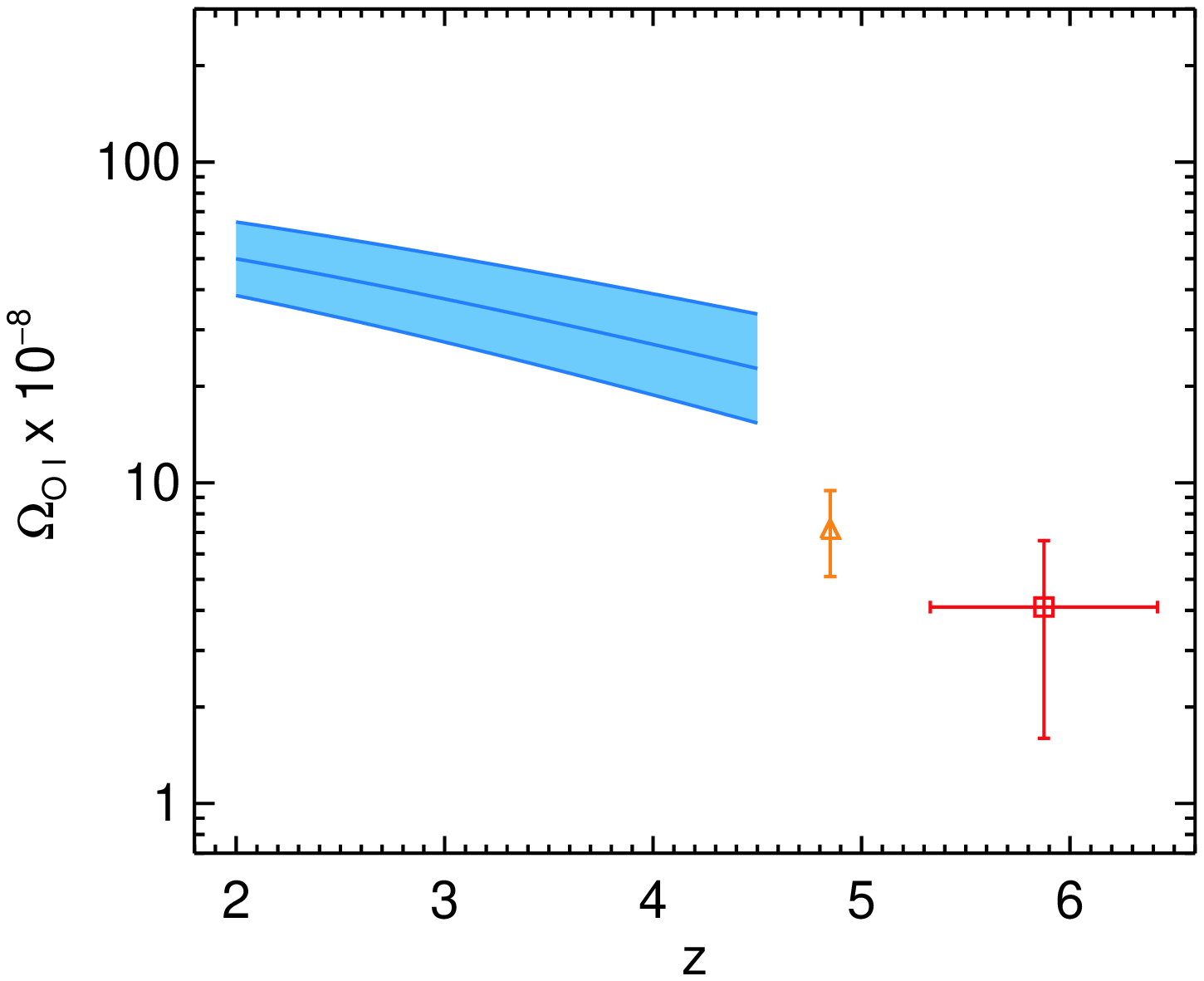}
\vspace{-5mm}
   \caption{Evolution of the comoving mass density of \oi, expressed
     as a fraction of the critical density.  The open square is
     derived from measurements of individual absorbers in
     \citet{becker2011b}.  The shaded region is an estimate based on
     fits to $\Omega_{\rm H\, I}(z)$ for DLAs from
     \citet{crighton2015} and the mean DLA metallicity from
     \citet{rafelski2014}.  The open triangle at $z=4.85$ is
     calculated from discrete $\Omega_{\rm H\, I}$ and mean DLA
     metallicity values from these works near $z \sim
     5$. \label{fig:Omega_OI}}
   \end{center}
\end{figure}

Low-ionization metal lines are potentially powerful probes of the
high-redshift Universe in multiple ways.  First, they trace the dense
gas in and around galaxies, offering insights into the interstellar
media of these objects analogous to the ways in which damped
\Lya absorbers (DLAs) trace the kinematics and composition of
lower-redshift galaxies \citep[e.g.][]{wolfe2005}.  Second, because
the metals in these systems are often dominated by a single ionization
state, determining abundances is relatively straightforward.  Finally,
if the last regions of the IGM to be reionized are metal enriched,
then they may give rise to a ``forest'' of absorption lines such as
\oi\ and \cii\ that can be detected in quasar spectra
\citep[e.g.][]{oh2002,furloeb2003}.

Metal-enriched gas where the hydrogen is largely neutral is traced by
lines such as \oi, \cii, \siii, and \feii.  \oi\ is particularly
useful for studying neutral gas since the first ionization potential
of oxygen and hydrogen are very similar, and due to charge exchange,
$n({\rm O}^{+})/n({\rm O}) \simeq n({\rm H}^{+})/n({\rm H})$ over a
wide range of physical conditions \citep[e.g.][]{osterbrock2006}.
Elements where the first ionization potential is significantly less
than 13.6 eV, such as carbon, silicon, and iron, are not shielded by
atomic hydrogen, and therefore appear as singly ionized species in
otherwise ``neutral'' gas.  At $z < 5$, low-ionization systems are
typically found by their strong, often damped, hydrogen \Lya lines.
At higher redshifts, however, the growing saturation of the
\Lya forest makes identifying individual \Lya absorbers difficult,
and metal systems must be identified using ``pseudo multiplets'' of
lines redward of Ly$\,\alpha$, such as \oi~$\lambda$1302;
\cii~$\lambda$1334; and \siii~$\lambda$1260, $\lambda$1304, and
$\lambda$1526.

\citet{becker2006,becker2011b} searched for low-ionization systems
over $5.3 < z_{\rm abs} < 6.4$.  In 17 lines of sight they find ten
systems with \cii\ and \siii, nine of which also contain \oi.
Infrared spectra have been used to obtain \feii\ for many of these
systems \citep{becker2012,dodorico2013}.  Among the nine \oi\ systems,
the ratios of \oi, \cii, \siii, and \feii\ column densities are
reasonably constant, which suggests that neither ionization
corrections nor dust depletion are large factors, as these would tend
to introduce scatter.  This supports the expectation that \oi\ systems
trace predominantly neutral gas, and are therefore the analogues of
lower-redshift damped \Lya systems (DLAs; $N_{\rm H\,I} \ge
10^{20.3}\rm\,cm^{-2}$ \citep[e.g.][]{wolfe2005} and sub-DLAs
($10^{19}{\rm \, cm^{-2}} < N_{\rm H\,I} < 10^{20.3}\rm\,cm^{-2}$)
\citep[e.g.][]{d-z2003}.  \footnote{Indeed, \citet{becker2011b} find that the metal line kinematics
of \oi\ systems at $z \sim 6$ are comparable to those of
lower-redshift DLAs and sub-DLAs, although the lines strengths are
weaker.}  It is therefore reasonable to compare
\oi\ systems at $z > 5$ to these systems, although it should be
emphasized that \hi-selected systems with $N_{\rm H\,I} > 10^{19}\rm\,cm^{-2}$
may not represent a complete census of \oi\ systems at lower
redshifts.

In terms of number density, \citet{becker2011b} find $dn/dX =
0.25^{+0.21}_{-0.13}$, which is similar to the combined number density
of DLAs and sub-DLAs over $3 < z < 5$
\citep{peroux2005,omeara2007,prochaska2009a,noterdaeme2009,guimaraes2009,crighton2015}.
At least one of the \oi\ systems lies close to the detection limit of
the existing data, however \citep{becker2011b}, so it is possible that
the number density of weak low-ionization systems (${N_{\rm O\,I}}
\lesssim 10^{13.5}\rm\,cm^{-2}$, $N_{\rm C\,II} \lesssim
10^{13.0}\rm\,cm^{-2}$) may be larger.  The observed mass density of
\oi\ in systems at $z \sim 6$ can be directly computed using
Eq.~(\ref{eq:omega_approx}) because the lines are typically
unsaturated, and so have measurable column densities.  At lower
redshifts \oi\ is generally saturated; however, the total mass density
can be estimated by combining measurements of $\Omega_{\rm H\,I}$ in
DLAs, which dominate the neutral gas budget
\citep[e.g.][]{noterdaeme2009,crighton2015} with the column
density-weighted mean DLA metallicity \citep[e.g.][]{rafelski2014}.
Fig.~\ref{fig:Omega_OI} shows $\Omega_{\rm O\,I}$ at $z \sim 6$
computed from the \citet{becker2011b} measurements, along with an
estimates over $2 < z < 4.5$ based on the fit to $\Omega_{\rm
  H\,I}(z)$ in \citet{crighton2015} and a fit to the mean metallicity
evolution at $z < 4.7$ from \citet{rafelski2014}.  A separate estimate
at $z=4.85$ is also given based on discrete $\Omega_{\rm H\,I}$ and
mean DLA metallicity measurements by these authors near $z \sim 5$.

While these mass density estimates should be treated with caution,
they broadly suggest a substantial buildup of metals in the low-ionization
phase over $4.5 \lesssim z \lesssim 6$, followed by a more gradual
buildup down to $z \sim 2$.  Given that $\Omega_{\rm H\, I}$ decreases
with time \citep{prochaska2009a,noterdaeme2009,crighton2015}, and that
the number density of low-ionization absorbers does not change rapidly
(see above), the increase in the mass density of low-ionization metals
may be largely driven by an increase in the (low-ionization phase) metal content of
the absorbers.  

\subsubsection{\mgii}

\begin{figure}
\begin{center}
\includegraphics[width=0.50\textwidth]{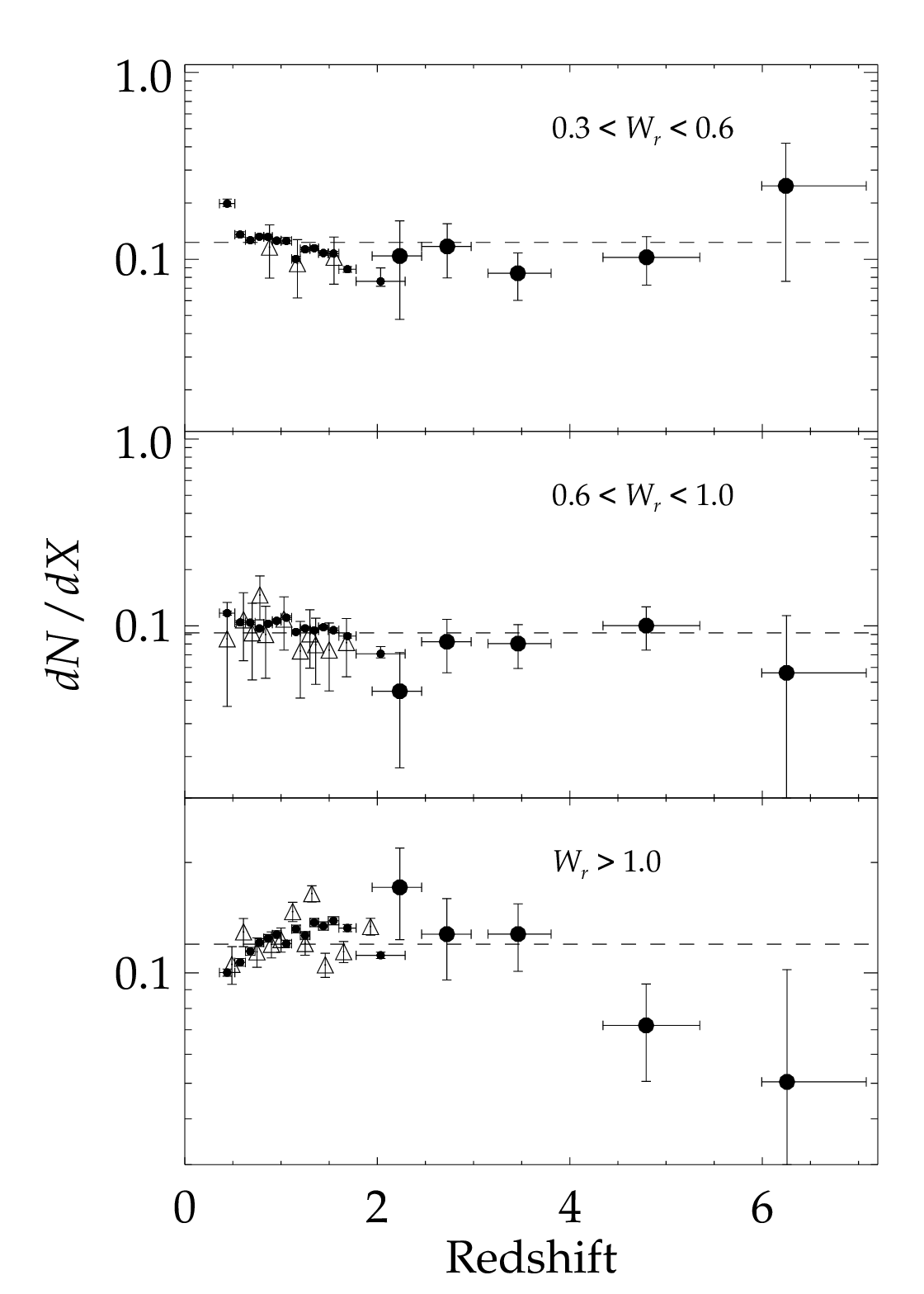}
\vspace{-4mm}
\caption{Evolution of the line-of-sight number density of
  \mgii\ absorbers.  The three panels give $dn/dX$ for three ranges of
  $\lambda$2796 rest-frame equivalent width.  Triangles are from
  \citet{nestor2005}.  Small circles are from \cite{seyffert2013}.  Large circles
  are from Chen et al. (in prep).  Dashed lines shows the mean $dn/dX$
  for the \citet{nestor2005} and \citet{seyffert2013} data.  For more details, see Chen et al. (in prep).  {\it Figure provided courtesy of R. Simcoe.  }}
\label{fig:MgII_incidence}
\end{center}
\end{figure}

The final absorption-line probe of metals at $z > 5$ is the
\mgii\ doublet, which can trace both neutral and ionized gas phases.
The first survey for \mgii\ over $2.5 < z < 6$ was conducted by
\citet{matejek2012}, followed by an expanded sample from Chen et al. (in prep), both using FIRE data.  Column
densities of strong \mgii\ systems are often difficult to measure due
to saturation effects, and it is therefore common to quantify these
systems in terms of the rest-frame equivalent width, $W_{\rm r}$, of
the 2796~\AA\ line.  Incidence rates of \mgii\ systems in different
ranges in $W_{\rm r}$ are shown in Fig.~\ref{fig:MgII_incidence}.  The
$z > 2$ data are compared to lower-redshift measurements using Sloan
Digital Sky Survey (SDSS) data from \citet{nestor2005} and
\citet{seyffert2013}.  For weak systems, the higher-resolution FIRE spectra tend to be more sensitive than the larger SDSS samples.  Incompleteness corrections are therefore important in
comparing these samples, particularly for $W_{\rm r} < 1$~\AA.  At
face value, however, the incidence rate of weak ($W_{\rm r} < 1$~\AA)
systems appears to relatively flat with redshift, while stronger
systems appear to become more numerous from $z \sim 6$ to $z \sim
2$-3, before declining towards lower redshifts.

The fact that the evolution in $dn/dX$ for strong \mgii\ systems shows
the same shape as the cosmic star formation rate density has been
cited as evidence that these systems are closely linked to
star-forming galaxies \citep{prochter2006,menard2011,matejek2012}.
\citet{menard2011} also note a correlation between $W_{\rm r}$ and the
      [O\,{\sc ii}] luminosity, a tracer of star formation, enclosed
      in an SDSS fiber.  By comparison, the relatively constant
      incidence of weak \mgii\ systems suggests that these absorbers
      are a feature of metal-enriched haloes that are established early
      in the process of galaxy formation, and change as a population
      relatively little, even as other galaxy properties evolve
      \citep[see discussion in][]{matejek2012}.

\subsubsection{Summary of observations: enrichment vs. ionization}

Although observations of metals near the reionization epoch are still
in an early phase, the results generally indicate substantial metal
enrichment in the interstellar and circumgalactic environments of
galaxies between $z \sim 6$ and 5.  The buildup of metals is expected
to correlate with the increase in stellar mass density; however,
changes in the ionization state of the metals must also play a role.
For example, the ratio of $N({\mbox{\siiv}})/N({\mbox{\civ}})$ in
\civ-selected systems tends to increase with redshift, an indication
that higher-redshift systems are tracing denser gas and/or a weaker
ionizing background \citep{dodorico2013,boksenberg2015}.  The fact
that the incidence of low-ionization systems remains roughly constant
out to $z \sim 6$, even while the number density of galaxies is
declining with redshift also suggests that the cross-sections of metal
enriched haloes that are largely neutral may be increasing with
redshift due to a declining UVB \citep{becker2011b,keating2014}.
\citet{matejek2013} also find evidence that \mgii\ systems tend to
trace DLAs, which are predominantly neutral, with increasing frequency
towards higher redshifts.  For high-ionization systems, shock heating may play a role in setting the ionization state \citep{cen2011}, along with the UVB.

\subsection{The metal mass budget}

We now examine in more detail what these observations tell us about
high-redshift galaxies and reionization.  First we ask whether the
observed mass density of metals at $z \sim 6$ is consistent with
expectations from the star-formation history at higher redshifts.
Following \citet{pettini1999}, we can estimate the global mass density
of metals by multiplying the stellar mass density by a nucleosynthetic
yield derived from stellar models \citep[for similar calculations at
  lower redshifts, see][]{peeples2014,shull2014}.  We emphasize that
there are considerable uncertainties in the stellar mass density at
these redshifts, the theoretical yields, and, as described above, the
measured metal mass densities.  This exercise is therefore intended
only to give a rough insight into whether the observed metals
constitute a reasonable fraction of the metals expected to be produced
at $z > 6$.

The stellar mass density at $z \sim 6$ for galaxies more massive than
$10^8~M_{\odot}$ is $\rho_{*} \sim 6 \times 10^6~{\rm
  M_{\odot}\,Mpc^{-3}}$ \citep{gonzalez2011}, taking into account the
correction factor of 1.6 suggested by \citet{stark2013} for
contamination due to nebular lines.  Following \citet{peeples2014}, we
adopt a total metal yield of $y = 0.030$ for Type II supernovae, which
should dominate the metal production at these early times.  The yields
of oxygen and carbon are taken as $y_{\rm O} = 0.015$ and $y_{\rm C} =
0.0083$, respectively.  Using these values, we would expect $\rho_{\rm
  O} \sim 9 \times 10^4~{\rm M_{\odot}\,Mpc^{-3}}$ and $\rho_{\rm C}
\sim 5 \times 10^4~{\rm M_{\odot}\,Mpc^{-3}}$, or $\Omega_{\rm O} \sim
6 \times 10^{-7}$ and $\Omega_{\rm C} \sim 4 \times 10^{-7}$.  By
comparison, the observed mass densities are $\Omega_{\rm O\,I} \sim 4
\times 10^{-8}$ \citep{becker2011b} and $\Omega_{\rm C\,II} +
\Omega_{\rm C\,IV} \sim 2 \times 10^{-8}$
\citep{rw2009,becker2011b,simcoe2011a,dodorico2013}.  The mass density
sampled via metal absorption lines is therefore only $\sim$5\% of the
expected total.  This indicates that the observed metals can easily be
produced by the known galaxies at these redshifts.  Furthermore, it
suggests that a large fraction of the metals reside in phases not
easily probed by absorption lines.  This includes the metals
re-accreted onto stars, in dense pockets of the interstellar medium
with small cross-sections, and in ionization states not directly
measured by the available lines.  These scenarios are seen in numerical simulations of metal enrichment \citep[e.g.,][]{oppenheimer2009,cen2011,finlator2015}
Similar conclusions have also been
reached at lower redshifts \citep[for a recent, more detailed
  accounting at $z \sim 0$ see][]{peeples2014}.

\subsection{Constraints on stellar populations}

\begin{figure}
\begin{center}
\includegraphics[width=0.4\textwidth]{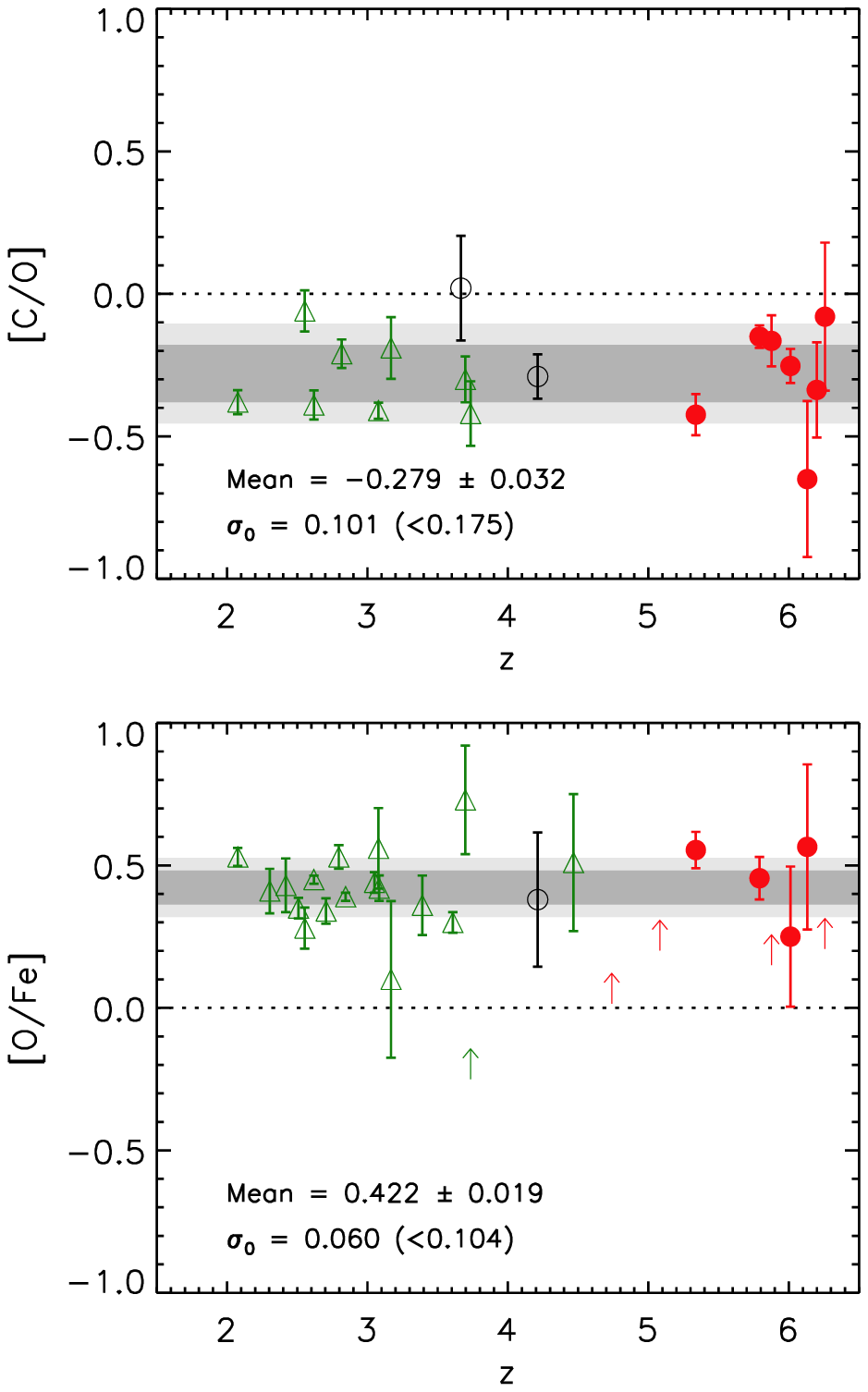}
\vspace{-4mm}
\caption{Relative abundances of carbon and oxygen (top panel), and
  oxygen and iron (bottom panel), expressed logarithmically as a
  fraction of the solar value.  Triangles and arrows at $z < 4.5$ are
  for metal-poor DLAs from \citet{cooke2011b}.  Open circles are for
  sub-DLAs from \citet{d-z2003} and \citet{peroux2007}.  Filled
  circles and arrows at $z > 4.7$ are for low-ionization absorbers
  from \citet{becker2012}.  The mean value given in each panel has
  been calculated from all measurements.  The nominal intrinsic rms
  scatter (accounting for measurement uncertainties), $\sigma_{0}$,
  and the 95\% upper limit on $\sigma_{0}$ (in parentheses) are shown
  as dark and light shaded bands, respectively.  {\it Reproduced from
    Figure 11 of \citet{becker2012} by permission of the authors and the AAS.}}
\label{fig:ratios}
\end{center}
\end{figure}

Although metal absorption lines provide only a tracer of the total
metal budget, their composition can nevertheless provide constraints
on the nature of the stars that formed during the reionization epoch.
At $z \gtrsim 5.5$ the \Lya forest becomes too thick to measure
\hi\ column densities of individual absorbers (\S\ref{sec:GP}), which
hinders direct metallicity measurements.  Relative abundances can
still be inferred, however, particularly for low-ionization systems
where ionization corrections should be minimal
\citep[e.g.][]{wolfe2005,becker2011b}.  The relative abundances of O,
C, Si, and Fe were measured for nine low-ionization systems at $z \sim
5$ to 6 by \citet{becker2012}.  Results for [C/O] and [O/Fe] are shown
in Fig.~\ref{fig:ratios}, where [X/Y] gives the logarithmic abundances
with respect to solar, ${\rm [X/Y]} = \log{(X/Y)} -
\log{(X/Y)_{\odot}}$.  The results at
$z \sim 5$-6 are generally consistent with relative abundances
measured in DLAs and sub-DLAs over $2 < z < 4$
\citep[e.g.][]{d-z2003,wolfe2005,peroux2007,cooke2011b}.  Moreover,
the absorption line measurements at high redshift are broadly
consistent with the abundances in (non carbon-enhanced) metal-poor
halo stars \citep[e.g.][]{cayrel2004}, particularly those with ${\rm
  [O/H]} \lesssim -2$ \citep{fabbian2009b} \citep[see discussions
  in][]{cooke2011b,becker2012}.  The [O/Fe] values are consistent with
enrichment from Type II supernovae \citep[e.g.][]{chieffi2004}, as
expected given there has been little time at $z \sim 5$-6 for a
contribution from Type Ia supernovae.  The lack of strong variations
in the absorption-line ratios suggests that these systems are enriched
by broadly similar stellar populations.  There is no clear evidence of
unusual abundance patterns that would indicate enrichment from exotic
sources such as Population III stars, although trends in [C/O] among
metal-poor DLAs and halo stars may indicate a Population III
contribution at the very low-metallicity end \citep[${\rm [O/H]} <
  -2$;][]{cooke2011b}.  In semi-analytic models of galaxy formation including metal enrichment, \citet{kulkarni2014} also conclude that measured ionic ratios at $z \lesssim 6$, particularly [O/Si], preclude a large contribution from Population III stars to the metal or ionizing photon budget during reionization.

\subsection{The connection to galaxies and reionization} 

Numerous studies over $0 < z < 5$ have demonstrated that the gas
traced by metal lines reflects a cycle of inflows and outflows that
help to regulate galaxy growth \citep[e.g.][]{dave2007,steidel2010,tumlinson2011,turner2014,crighton2015}.  The lower-redshift studies have shown that
galaxies over a wide range of luminosities and star-formation rates
are surrounded by an enriched circumgalactic medium out to typical
distances of at least $\sim$100 proper kpc
\citep[e.g.][]{simcoe2006a,bordoloi2014}, indicative of the volume
filled by outflows or other enrichment mechanisms.  Metal lines at $z
\sim 5-6$ should therefore trace the early stages of the enrichment
process, helping to elucidate the mechanisms that shape the earliest
galaxies.  In addition, if metal absorption lines tend to be
associated with low-mass galaxies, then these lines may help to
identify vital sources of ionizing photons that are beyond the reach
of current direct galaxy surveys.

\begin{figure*}
   \centering
   \begin{minipage}{\textwidth}
   \begin{center}
      \includegraphics[width=0.6\textwidth]{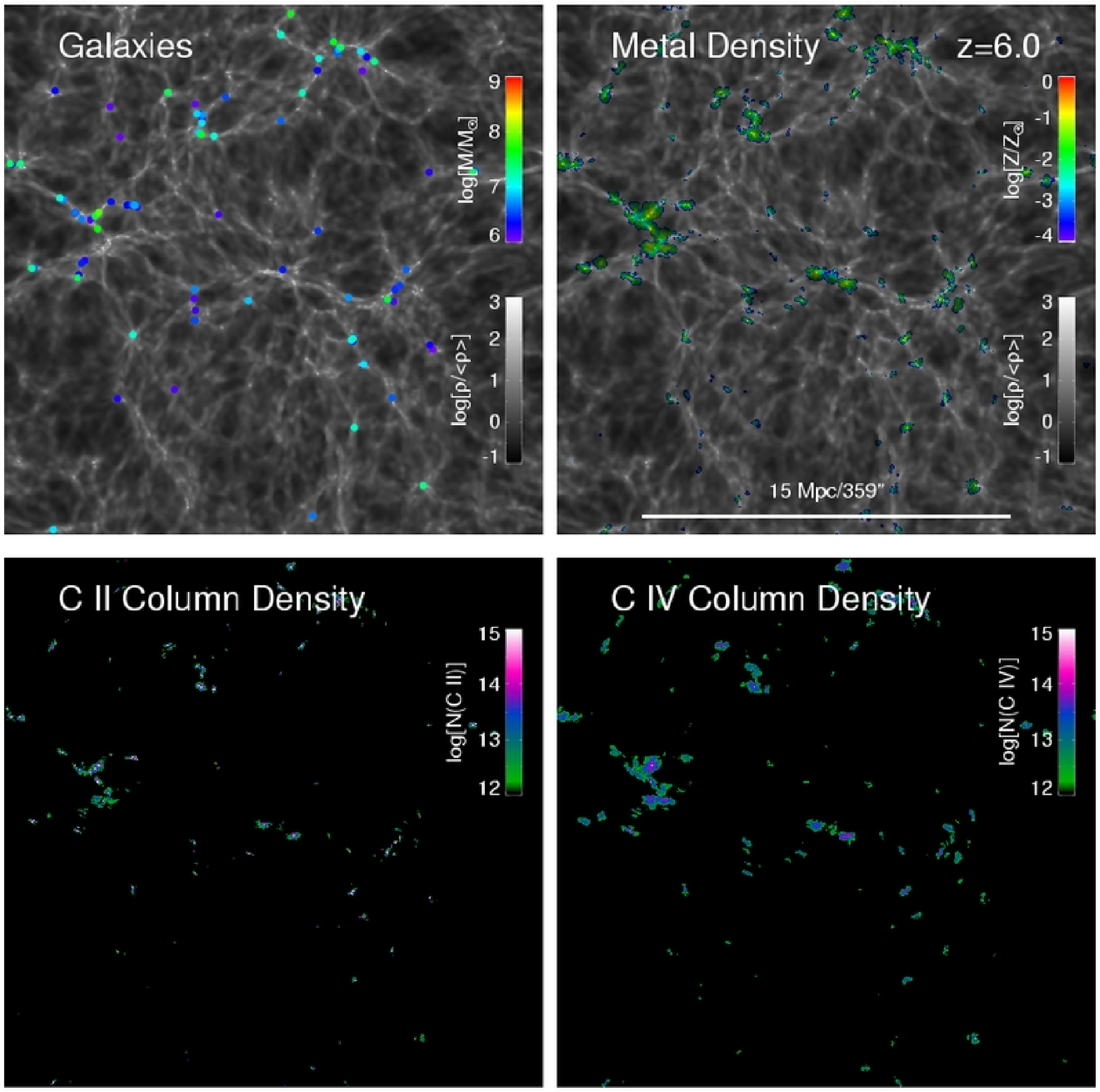}
      \vspace{-0.2cm}
      \caption{A slice through a hydrodynamical simulation at $z=6.0$
        showing the location of galaxies with stellar masses greater
        than $10^{6}~{\rm M}_{\odot}$ (upper left), the metallicity as
        a fraction of solar (upper right), and the integrated
        \cii\ (lower left) and \civ\ (lower right) column densities.
        The grey scale in the upper panels shows the total mass
        density as a fraction of the mean density.  The slice is
        16\,$h^{-1}$\,Mpc on a side and 25 km\,s$^{-1}$ thick.  In this simulation, metal
        absorbers closely trace the galaxies, with only a small
        fraction of the volume enriched above $1/10,000^{\rm th}$ of
        solar metallicity.  The cross-section of \civ, which favors
        lower densities, is more extended than that of \cii.  {\it
          Reproduced from Figure 5 of \citet{oppenheimer2009} by permission of the
          authors.}}
   \label{fig:metal_map}
   \end{center}
   \end{minipage}
\end{figure*}

We can use lower-redshift observations to gain some insight into the
type of galaxies associated with metal absorbers near $z \sim 6$.  The
line-of-sight number density of \civ\ absorbers at $z \sim 6$ with
$N_{\rm C\,IV} > 10^{13}~{\rm cm^{-2}}$ is $dn/dX \sim 1$
\citep{dodorico2013}.  For a population with a fixed physical
cross-section, $\sigma$, the comoving number density can be computed
as
\begin{equation}
   \label{eq:number_density}
   \phi = \frac{dn}{dX}\frac{H_{\rm 0}}{c\,\sigma} \, .
\end{equation}
If we take the lower-redshift value of 100 kpc (proper) as an upper
limit on the radius to which haloes are enriched at $z \sim 6$, then
the number density of such haloes would be $\phi \sim 7 \times
10^{-3}~{\rm Mpc}^{-3}$.  This corresponds to the number density of
galaxies with absolute UV magnitudes $M_{\rm UV} \lesssim -17$ to $-$18
\citep{Finkelstein2014,Bouwens2015},
which is near the limit of direct galaxy surveys.  In terms of dark
matter haloes, this number density corresponds to haloes with masses
$M_{\rm h} \gtrsim 2 \times 10^{10}~{\rm M_{\odot}}$
\citep{murray2013}.  These figures already suggest that metal
absorbers at $z \sim 6$ are likely to be associated with relatively
modest galaxies, particularly given that an enrichment radius of 100
kpc is probably a conservative upper limit due to the limited time
available for metal-enriched outflows to travel by $z \sim 6$
\citep[e.g.][]{oppenheimer2009}.  Further insights can be gained by
directly searching for galaxies associated with absorbers, and through
more sophisticated modelling.

Directly identifying the galaxies associated with metal absorbers at
these redshifts is challenging due to the faintness of the sources and
the corresponding difficulty in obtaining spectroscopic redshifts.
\citet{diaz2011,diaz2014,diaz2015} have searched for galaxies
associated with \civ\ systems at $z \simeq 5.7$.  In at least one
case, on $\sim$10\,$h^{-1}\,{\rm Mpc}$ scales the \civ\ absorber
appears to be more closely associated with narrow band-selected
\Lya emitting galaxies (LAEs), rather than broad band-selected Lyman
break galaxies (LBGs).  This suggests that \civ\ systems may trace
lower-density environments dominated by low-mass galaxies
\citep{diaz2014}.  Further efforts at identifying galaxies associated
with \civ\ and other types of metal absorbers will help to clarify
this picture.

On the theoretical side, considerable efforts have been made to model
the production and distribution of metals near reionization using
hydrodynamical simulations
\citep{oppenheimer2006,oppenheimer2008,dave2007,oppenheimer2009,cen2011,keating2014,pallottini2014,finlator2015}
and analytical methods \citep{bagla2009}.  The simulations generally
support a picture where metals reside in overdense regions, and that
only a few percent of the IGM needs to be enriched in order to
reproduce the observed line statistics (Fig.~\ref{fig:metal_map}).
Metal lines should therefore closely trace the environments of
galaxies, and be sensitive to the details of feedback mechanisms
\citep[e.g.][]{oppenheimer2006,oppenheimer2008,cen2011}.
\citet{finlator2015} further showed that ion ratios are sensitive to
the local variations in the UVB induced by these galaxies \citep[see also][]{oppenheimer2009}.

A further conclusion of simulations is that, while absorbers arise
from galaxies over a wide range in mass, low-ionization lines tend to
trace lower-mass galaxies than
\civ\ \citep[e.g.][]{oppenheimer2009,finlator2013}.  This occurs
because strong outflows are required to transport enriched material
out to the low densities where \civ\ becomes a favored ionization
state.  \citet{oppenheimer2009} find that \civ\ absorbers with $N_{\rm
  C\,IV} \ge 10^{13}~{\rm cm}^{-2}$ should largely be found around
galaxies with stellar masses $M_{*} = 10^{6.5-8.5} \rm\,M_{\odot}$
(UV magnitudes $M_{\rm UV} \simeq -14.5$
to $-$19), whereas \cii\ absorbers should trace galaxies that are
roughly a factor of 10 less massive.  In both cases the absorber
strength is found to anti-correlate with projected distance from the
galaxy, and for \civ\ it is found to correlate with galaxy stellar
mass.  The likelihood that metal lines, and in particular
low-ionization lines, trace low-mass galaxies underlines the
importance of these lines as probes of the faint galaxies that may be
largely responsible for reionization.

\begin{figure}
\begin{center}
\includegraphics[width=0.45\textwidth]{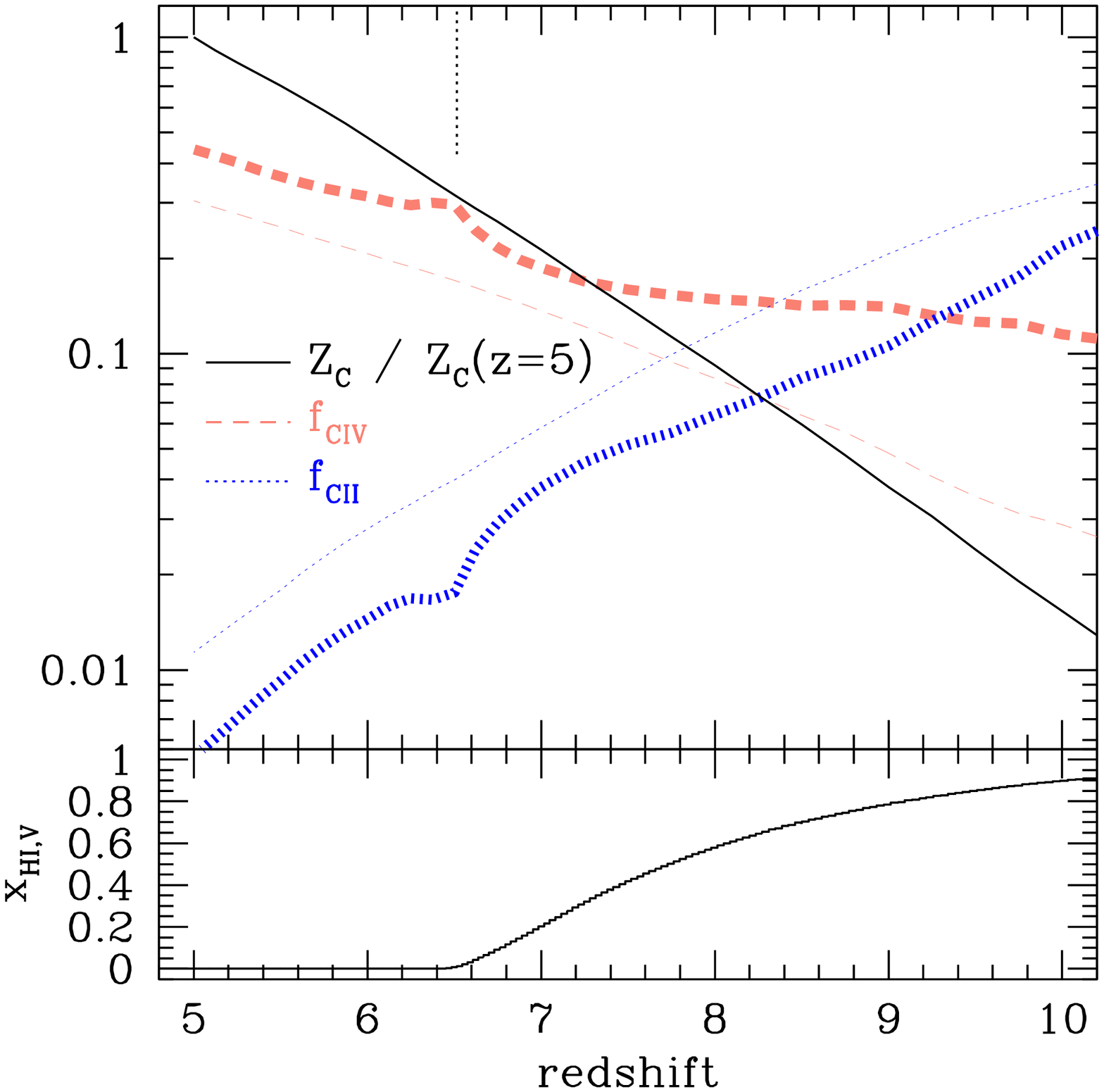}
\caption{Evolution of the carbon (top panel) and hydrogen (bottom
  panel) in a radiation hydrodynamical simulation by
  \citet{finlator2015}.  The solid line in the upper panel shows the
  total metal mass fraction, normalized to its value at $z = 5$.  The
  red dashed and blue dotted lines show the volume-weighted fractions
  of carbon that are in \civ\ and \cii, respectively.  The thick and
  thin lines give the results for the UV background predicted by the
  simulation and for a uniform \citet{HaardtMadau2012} background.  The solid
  line in the bottom panel shows the volume-weighted fraction of
  neutral hydrogen.  Simulations similar to this one predict that
  low-ionization metal lines should become more prominent relative to
  high-ionization lines in the reionization epoch.  {\it Reproduced
    from Figure 6 of \citet{finlator2015} by permission of the authors.}}
\label{fig:carbon_frac}
\end{center}
\end{figure}

The hydrodynamical simulations discussed above demonstrate that the
redshift evolution of essentially all ions is sensitive to (i) the
increase in the total mass density of metals towards lower redshifts
due to ongoing star formation, (ii) the propagation of enrichment
towards lower densities with declining redshift, and (iii) the decline
in the mean UV background with redshift at $z > 5$ (see also
\S\ref{sec:PIrate}).  The last two factors mean that systems traced by
\civ\ (\cii) should constitute a decreasing (increasing) fraction of
the metals towards higher redshifts (Fig.~\ref{fig:carbon_frac}).
This helps to explain the substantial decline in \civ\ from $z \sim 5$
to 6, even while the number density of low-ionization absorbers
remains relatively constant
\citep{oppenheimer2009,keating2014,finlator2015}.  If these trends
continue, then low-ionization lines may become numerous at $z \gtrsim
7$ \citep{finlator2015}.  Some authors have suggested that a
``forest'' of low-ionization lines such as \oi\ and \cii\ may appear
during reionization if significant amounts of the IGM are enriched at
earlier times \citep{oh2002,furloeb2003}.  At present there is
little evidence for an \oi\ forest in the $z \sim 6$ data; however,
higher-redshift lines-of-sight may yet reveal substantial quantities of
enriched, neutral gas in the reionization epoch.  The prospects for this depend partly on whether
early metal enrichment extends to truly intergalactic regions, or if it is mainly confined
to the circum-galactic environments around galaxies.

\section{THE REIONIZATION HISTORY}
\label{sec:reion_mods}

Thus far we have examined how quasar absorption line observations
elucidate the properties of high-redshift galaxies by probing the
post-reionization UVB and heavy element production in the early
Universe.  In this section, we now turn to focus on how quasar
absorption line studies directly constrain the reionization history at
$z\gtrsim 6$.  We begin by briefly describing the basic properties of the
EoR relevant for current and future observations of the IGM, and then
proceed to examine the current quasar absorption line data.

\subsection{The IGM during reionization}

\begin{figure}
\begin{center}
\includegraphics[width=\columnwidth]{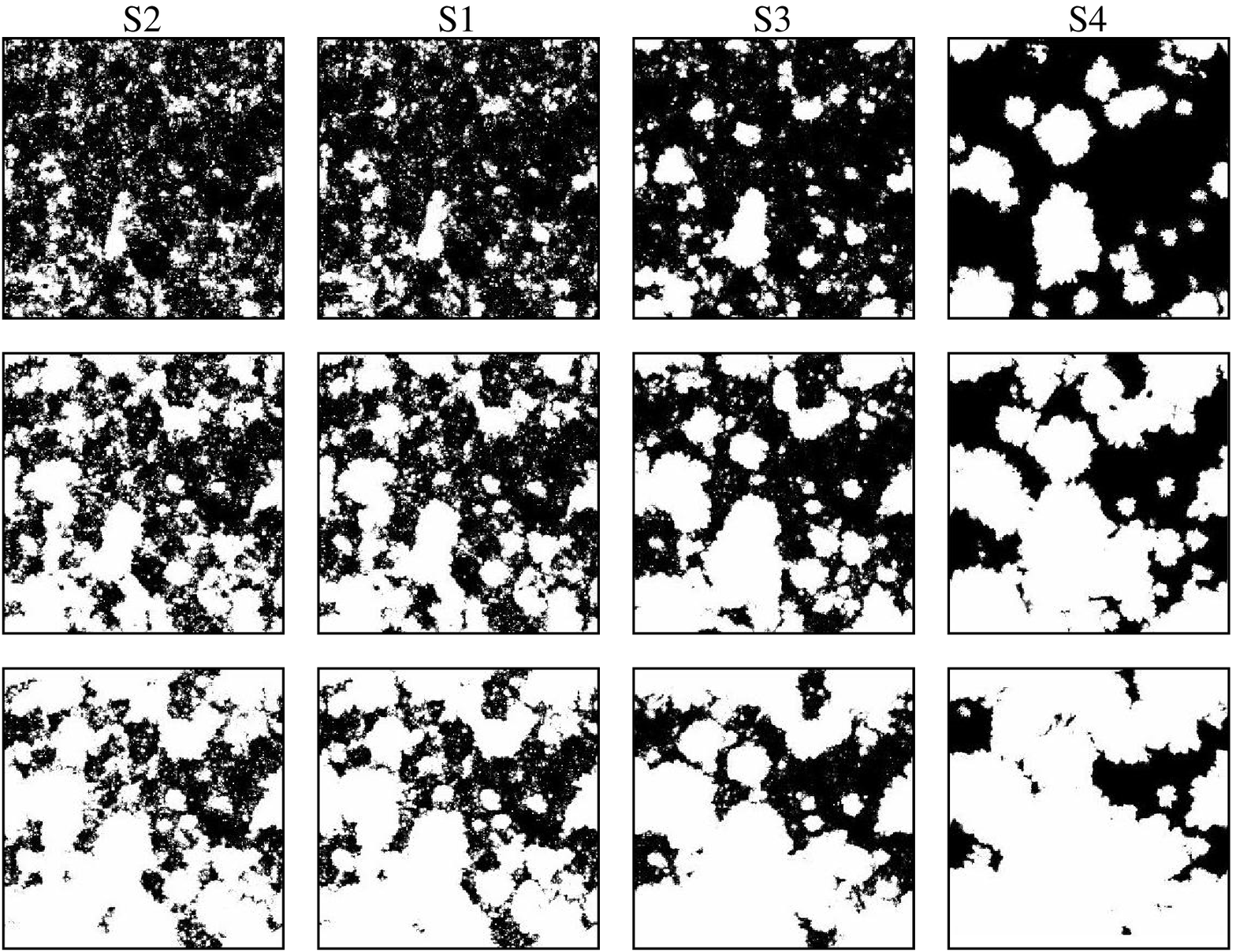}
\caption{Models illustrating the size and distribution of ionized
  regions at different stages of reionization. Each panel shows a
  slice ($0.25 h^{-1}$ comoving Mpc thick) through a numerical
  simulation of reionization, spanning $65.6h^{-1}$ comoving Mpc on a
  side. The white regions show highly ionized hydrogen while the dark
  regions are neutral.  The columns, from left to right, show
  different models for the ionizing sources which reside in
  progressively more massive and highly clustered host haloes. The
  rows, from top to bottom, show different stages of the reionization
  process: in each model the efficiency of the ionizing sources is
  normalized so that the mean ionized fraction is $\avg{x_i} = 0.2,
  0.5$, and $0.7$, respectively.  {\it Reproduced from
    Figure 3 of \citet{McQuinn2007} by permission of the authors.} }
\label{fig:bubbles}
\end{center}
\end{figure}

The IGM is expected to resemble a two-phase medium during
reionization, with part of the IGM in highly ionized ``bubbles'' that
form around collections of galaxies and accreting black holes, while
the rest remains mostly neutral (e.g. Fig. \ref{fig:bubbles}).  This
two-phase medium can be broadly characterized by two quantities: the
volume filling factor of ionized hydrogen and the size distribution of
the ionized regions \citep[see
  e.g][]{Madau1999,MiraldaEscude2000,Gnedin2000,Furlanetto2004nh}.  A primary goal of
reionization studies is to constrain these properties from
observations, and to then use them to inform models of galaxy
formation and the high-redshift IGM.

However, this two-phase description does not completely describe the
ionization state of the IGM.  Firstly, there is also some residual
neutral hydrogen within the bubble interiors; this neutral fraction
will shrink during and after reionization as the mean free path of the
ionizing photons grows; this is eventually observed as the \Lya forest
at $z<6$. Secondly, in addition to this highly ionized diffuse gas,
there will be some mostly neutral, self-shielded dense clumps (LLSs)
that remain, both in the bubble interiors and in the post-reionization
IGM \citep{FurlanettoOh2005,GnedinFan2006,Choudhury2009,AlvarezAbel2012,SobacchiMesinger2014}.
These systems play an important role in setting the mean free path for ionizing photons.
Consequently, at the tail-end of reionization the IGM can be almost
completely filled by ionized gas once bubbles have overlapped, yet the
mean free path of ionizing photons nevertheless has significant
spatial fluctuations (\S \ref{sec:UVfluc}).  This intermediate period
is what \citet{FurlanettoOh2005} refer to as the transition between
the ``bubble-dominated and cosmic-web dominated'' eras in the
reionization history of the Universe. Capturing this transition era is
challenging for modellers, in part because of the large dynamic range
in spatial scale involved: this requires fully resolving the dense
sinks of ionizing photons that (mostly) regulate the mean free path of
the ionizing photons at the end of reionization, while simultaneously
capturing representative samples of the ionized bubbles and the large-scale spatial variations in the source abundance.

Quasar absorption line measurements play an important role within this
context.  First, these observations presently provide our most
detailed probe of the properties of the high-redshift IGM, with
current studies extending out to $z \sim 7$
\citep{Fan2006,Mortlock2011}.  In terms of constraining the redshift
evolution of the filling factor of ionized regions, these observations
may then address whether reionization completes at $z \leq 7$ or at
some earlier time.  As we discuss below, there are some interesting
hints that reionization may be incomplete at $z \leq 7$, and there are
good prospects for placing more definitive constraints in the future.

\subsection{Mean \Lya forest transmission}
\label{sec:meanf_a}

\begin{figure}
\begin{center}
\includegraphics[width=0.48\textwidth]{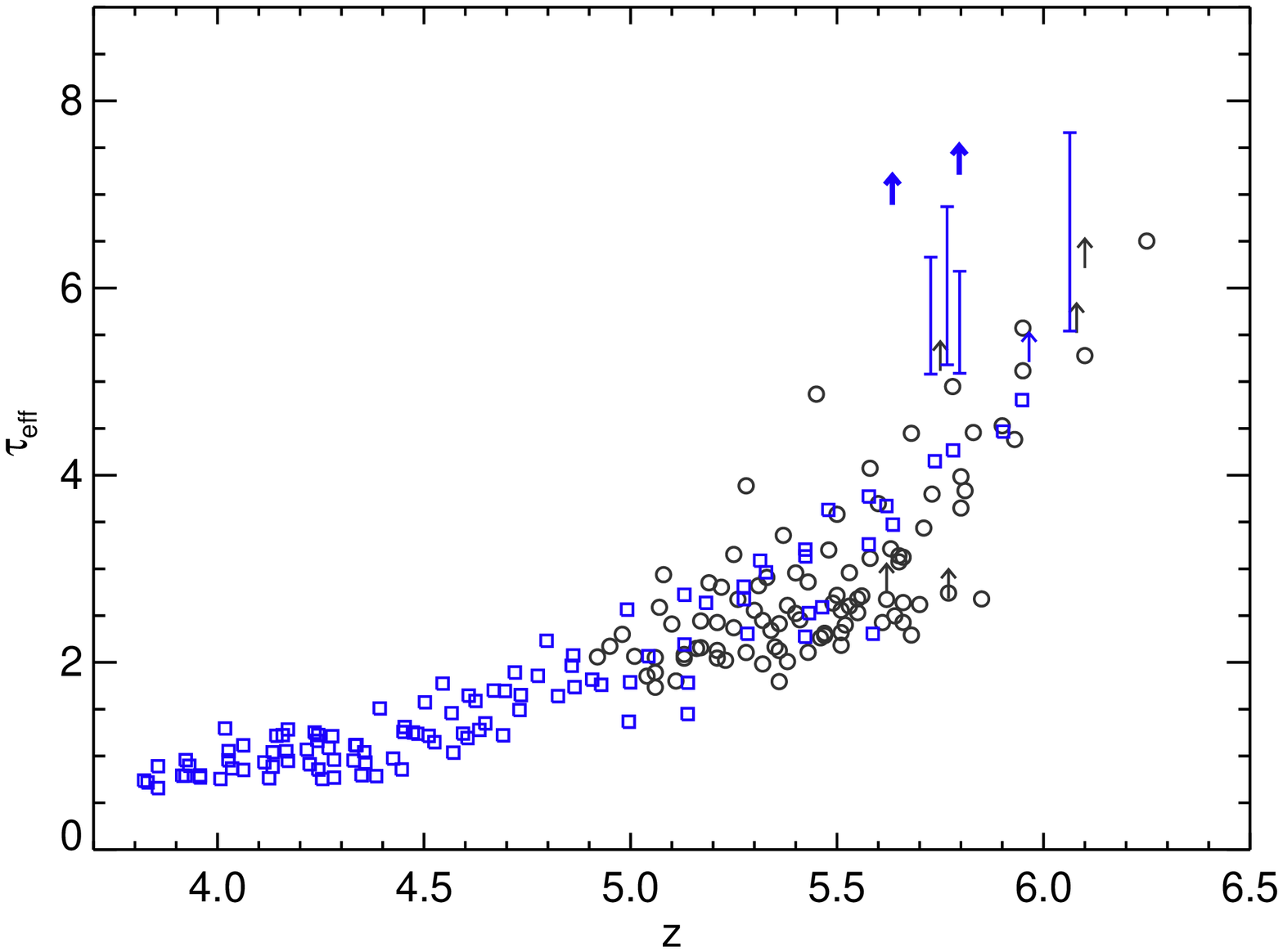}
\vspace{-0.7cm}
\caption{The evolution of the \Lya forest effective optical depth with
  redshift.  The measurements are from \citet{Fan2006} (black circles
  and arrows, $\Delta z = 0.15$ bins) and \citet{Becker2015} (blue
  circles, error bars and arrows, $\Delta l = 50h^{-1}$ Mpc bins).
  The lower limits at $z>5.5$ are obtained where the \Lya absorption
  saturates in the presence of a Gunn-Peterson trough. {\it Reproduced
    from Figure 6 of \citet{Becker2015} by permission of the authors.}}
\label{fig:tauGP}
\end{center}
\end{figure}

The first measurement to consider is the redshift evolution of the
average \Lya forest transmission as a function of redshift, $\langle F
\rangle$.  As is evident from the \cite{GunnPeterson1965} argument
presented in \S\ref{sec:GP}, at the redshifts of interest even gas
with neutral fractions of $\avg{x_{\rm HI}} \gtrsim
10^{-4}$--$10^{-5}$ should produce highly saturated absorption in the
\Lya line.  Although this large opacity makes direct inferences about
reionization from the \Lya forest challenging, progress can still be
made.  Fig.~\ref{fig:tauGP} displays the observed redshift evolution
of the effective optical depth, $\tau_{\rm eff}^{\alpha}=-\ln \langle
F \rangle$, in the \Lya forest at $z>3.8$.  Each point shows an
estimate from an individual stretch of spectrum of width either
$\Delta z \sim 0.15$ or $50h^{-1}$ comoving Mpc, which are similar at
the redshifts of interest.  The upward pointing arrows show 1-$\sigma$
lower bounds on the effective optical depth in regions where the
average transmitted flux is consistent with zero.

Fig.~\ref{fig:tauGP} reveals several striking features.  Firstly, the
presence of any transmission at $z \leq 6$ might seem to require that
reionization completed by $z=6$.  However, this is not completely
secure \citep{Lidz2007,Mesinger2010}. The transmission only demands
that some regions in the IGM are highly ionized: since the forest
shows a mixture of completely opaque and partly transmissive regions
at $z =5-6$ \citep{Fan2006}, it may be that some of the opaque regions
are actually significantly neutral, with order unity neutral
fractions. It appears feasible that some so-called ``neutral islands'' remain as
late as $z \sim 5.5$ \citep{MalloyLidz2015}; this possibility is
constrained by the dark pixel fraction tests described in \S
\ref{sec:dark_frac}.

Additionally, the redshift evolution of $\tau_{\rm eff}^{\alpha}$ near
$z \sim 6$ may be interpreted as being extremely rapid, requiring
reionization to complete near $z \sim 6$
(e.g. \citealt{Fan2006,GnedinFan2006}). However, it remains unclear
whether the mean transmission should evolve rapidly as reionization
completes. This behaviour is predicted in numerical simulations if the
mean free path, and hence photo-ionization rate incident on a typical
region, grows rapidly as reionization finishes
(\citealt{Gnedin2000,Bauer2015,Chardin2015}). This may not happen,
however, if the mean free path to ionizing photons at the end of
reionization is limited by dense photon sinks, and not the size of the
ionized regions themselves
\citep{FurlanettoMesinger2009,AlvarezAbel2012}. 

Finally, the small levels of remaining transmission through the \Lya
forest near $z \sim 5$--$6$ trace only rare, underdense regions of the
IGM \citep{BoltonBecker2009}. This is the case even if reionization is
complete at these redshifts, and the IGM gas is highly ionized.
Extrapolating from the behaviour in low-density voids to make
inferences about the overall ionization state in more typical regions
of the IGM can therefore be uncertain and model dependent
\citep{OhFurlanetto2005}.  In summary, while the mean \Lya forest
transmission clearly indicates the IGM neutral fraction is increasing
toward $z=6$, on its own it is unable to unambiguously determine the
duration and end-point of the reionization process.

\subsection{Higher-order Lyman series lines}
\label{sec:higher_order}

Further progress can be made by considering the transmission through
higher-order Lyman-series lines. These transitions have smaller
cross-sections\footnote{For Ly$\,\beta$, the rest-frame wavelength of
  the transition is $\lambda_{\rm Ly\beta}=1025.72\rm~\AA$ and the
  scattering cross-section is $\sigma_{\rm Ly\beta}=7.18\times
  10^{-19}\rm\,cm^{2}$.  Similarly for Ly$\,\gamma$, $\lambda_{\rm
    Ly\gamma}=972.54\rm~\AA$, $\sigma_{\rm Ly\gamma}=2.50\times
  10^{-19} \rm\,cm^{2}$.} relative to Ly$\alpha$, and so become fully
saturated at larger neutral hydrogen fractions.  The analogue of the
Gunn-Peterson formula (Eq.~\ref{eq:GP}) for \Lyb is then simply: \beq
\tau^\beta_{\rm GP} = \frac{\sigma_\beta}{\sigma_\alpha}
\tau^\alpha_{\rm GP} = 0.16 \tau^\alpha_{\rm GP}.
\label{eq:tau_gpb}
\eeq It is important to keep in mind, however, that it is the
effective optical depths, $\tau_{\rm eff}=\langle F \rangle$, which
are observable, and these do not scale in the same way as the true
optical depths
(e.g. \citealt{SongailaCowie2002,OhFurlanetto2005}). The ratio of
$\tau_{\rm eff,\alpha}/\tau_{\rm eff,\beta}$ depends on the gas
density distribution, the relation between temperature and density
(\citealt{FurlanettoOh2009}), and other aspects of the IGM model.  In
typical models considered in the literature, $\tau_{\rm eff,
  \alpha}/\tau_{\rm eff, \beta} \sim 3$ (e.g. \citealt{Fan2006}),
which is around a factor of two smaller than the ratio of the \Lya and
\Lyb cross-sections. The corresponding factors for higher-order lines,
such as Ly$\gamma$, are still larger.

One complication here is that the higher-order transitions land in
regions of the spectrum that also contain absorption from lower-order
lines, sourced by gas at lower redshift
(e.g. Fig.~\ref{fig:qso_examp}).  For example, a wavelength that
contains \Lyb absorption at $z_\beta = 6$ also contains \Lya
absorption from gas at $z_\alpha = \lambda_\beta
(1+z_\beta)/\lambda_\alpha -1 = 4.9$. However, since these two sources
of absorption are widely separated in physical space, they are to a
very good approximation uncorrelated \citep{Dijkstra2004Lyb}, and so:
\beq \avg{e^{-\tau_{\rm tot}}(\lambda)} =
\avg{e^{-\tau_\beta}(z_\beta)} \avg{e^{-\tau_\alpha}(z_\alpha)}.
\label{eq:amf_tot}
\eeq The average transmission through the \Lya line at $z_\alpha$ can
be easily measured using lower-redshift quasar spectra.  This
expression further generalizes to the \Lyg region of the quasar
spectrum -- which contains overlapping absorption from \Lyg and
lower-redshift absorption in \Lyb and \Lya -- and to
still higher-order lines.

In practice, measurements of the mean transmission in the \Lyb
and \Lyg forest -- after dividing out estimates of the
foreground absorption in the lower-order lines as described above --
are also mostly consistent with saturated absorption above $z \gtrsim
6$ \citep{Fan2006}. This strengthens the case that the opacity is
evolving near $z \sim 6$, and allows a slightly stronger limit on the
possibility that the absorption arises from highly ionized gas after
reionization.  Nevertheless, even the current upper limits on the
transmission through the \Lyg forest imply fairly modest lower
limits on the neutral fraction of $\avg{x_{\rm HI}} \gtrsim 10^{-4} -
10^{-3}$, depending on the precise model.  Hence, even high effective optical depth
absorption in \Lyg does not, by itself, imply that reionization
is incomplete at $z \gtrsim 6$.  However, higher-order Lyman series
lines are also useful for constraining the possibility that there are
neutral islands left over at $z \leq 6$, since these lines help in
placing upper limits on the dark pixel fraction, as described next.

\subsection{Dark pixel fraction and dark gaps}
\label{sec:dark_frac}

Recent work has demonstrated that a conservative, almost model-independent
lower bound on the filling factor of ionized regions can be placed by
counting the fraction of spectral pixels that are completely absorbed
\citep{McGreer2011,McGreer2015}.  This is because regions with
transmission through the \Lya line are certain to contain highly
ionized gas.  One can place a stronger limit by including the
\Lyb region of the quasar spectrum and potentially other higher
series lines. In this sense, the dark pixel fraction analysis directly
targets one of the main properties of the EoR, the filling factor of ionized bubbles.  However, the trade-off
is that the limit obtained is conservative and highly-ionized regions
may be responsible for some (or all) of the dark pixels.

In the first analysis of this sort, \citet{McGreer2011} found that the
volume-averaged neutral fraction needs to be smaller than $\avg{x_{\rm
    HI}} < 0.2$ at $5 \leq z \leq 5.5$, and smaller than $\avg{x_{\rm
    HI} }< 0.5$ at $z=6$, each at 1-$\sigma$ confidence. A subsequent
study incorporating new, higher signal to noise spectra has improved
these limits to $\avg{x_{\rm HI}} < 0.11$ at $z=5.6$, $\avg{x_{\rm
    HI}} < 0.09$ at $z=5.9$ at 1-$\sigma$ confidence
\citep{McGreer2015}.  At $z=6.1$, there are significantly more dark
pixels and the limit is weaker with $\avg{x_{\rm HI}} < 0.58$, again
at 1-$\sigma$ confidence.  These results suggest that if significantly neutral
diffuse gas remains at $z \leq 6$, it fills a rather small fraction of the IGM volume.

An approach to search for direct signatures of any remaining neutral islands 
is to stack spectra around long,
highly absorbed regions in the \Lya and \Lyb forests \citep{MalloyLidz2015}. If
significantly neutral gas remains in some of the fully absorbed
regions, the stacked spectra should recover slowly as one moves from
absorbed to transmitted regions, owing to the damping wing of the \Lya
line from diffuse neutral gas in the IGM
\citep{MiraldaEscude1998,MalloyLidz2015}.  However, one can only apply
this test at redshifts where the spectra show both completely absorbed
regions and regions with some transmission. In typical regions of the
IGM, this means this test can only be applied at $z \lesssim 6$.

\begin{figure}
\begin{center}

\includegraphics[trim=0cm 0cm 2.2cm 5.8cm, clip=true, width=\columnwidth]{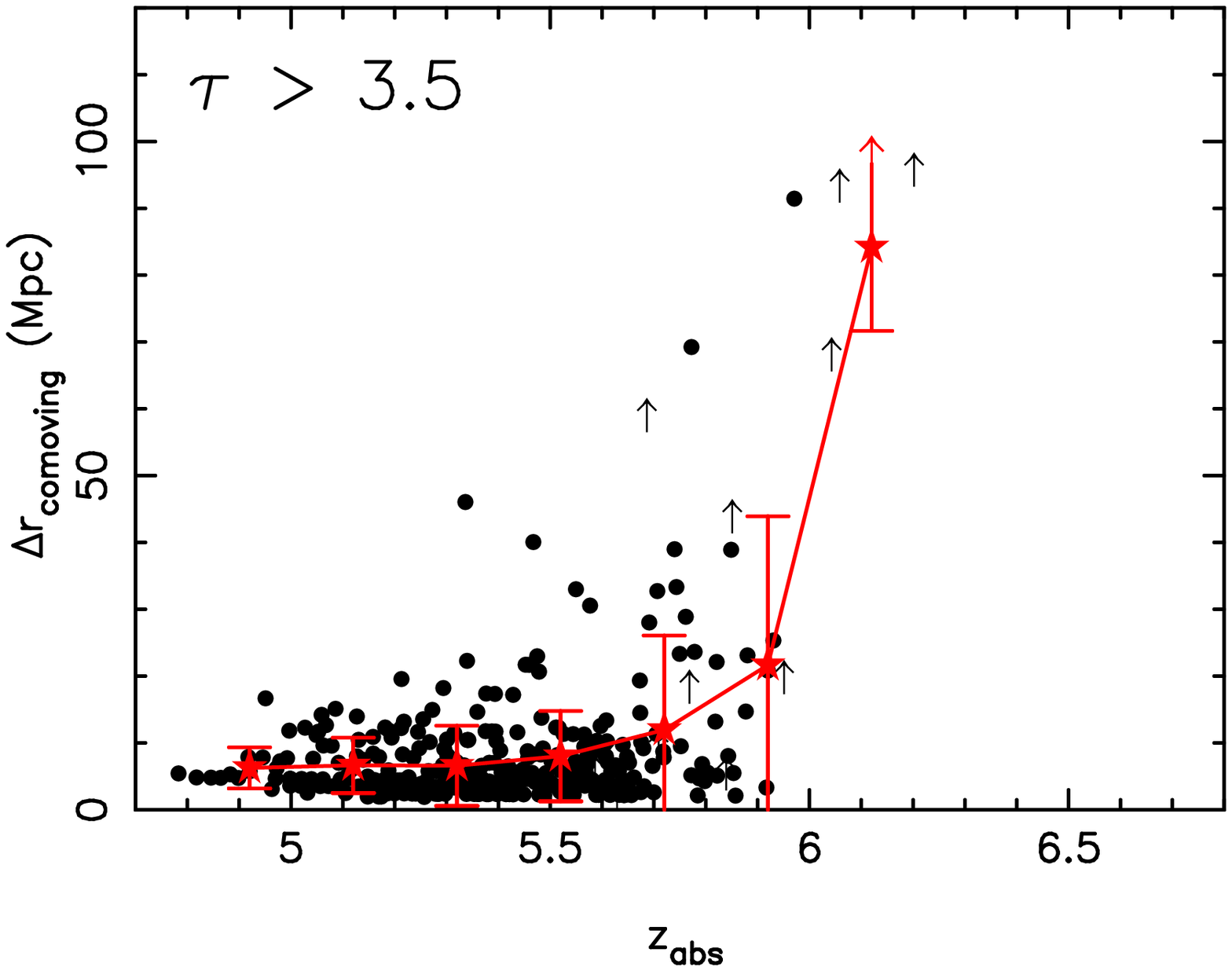}
\vspace{-0.8cm}
\caption{The size of dark gaps as a function of redshift measured from
  12 quasar spectra obtained with Keck/Echellete Spectrograph
    and Imager.  The gaps are defined as contiguous regions in the
  observed \Lya forest where the optical depth never goes beneath the
  threshold of $\tau_{\alpha} < 3.5$. The upward pointing arrows show
  lower limits on the dark gap size for observed gaps that are close
  to the quasar near-zone (see \S \ref{sec:prox}).  The spectra at $z
  \gtrsim 5.2$ show fully absorbed regions that are tens of comoving
  Mpc in size, and the dark gap sizes grow steeply in length at $z
  \gtrsim 5.7$. {\it Reproduced from Figure 10 of \citet{Fan2006} by permission of
    the authors and the AAS.}}
\label{fig:fan_gaps}
\end{center}
\end{figure}

Finally, a related diagnostic is the size distribution of contiguous
saturated regions in the \Lya forest, and the redshift evolution of
these so-called ``dark gaps''
(e.g. \citealt{SongailaCowie2002,PaschosNorman2005,Fan2006,Gallerani2006,Gallerani2008,Mesinger2010}). Some
measurements of dark gap sizes and their redshift evolution are shown
in Fig. \ref{fig:fan_gaps}. As with many of the other observed
properties of the $z \sim 6$ \Lya forest, the abundance of large dark
gaps grows steeply near $z \sim 6$. On the other hand, even by $z \sim
5.3$, there are some $\gtrsim 30$ Mpc contiguous regions that appear
entirely absorbed, with $\tau_{\alpha} \geq 3.5$ across the entire
stretch of spectrum.  This evolution might result in part from the
presence of remaining neutral islands in the IGM, or may instead
reflect the thickening of the \Lya forest owing to the increasing mean
density of the Universe and the dropping intensity of the UVB, along
with fluctuations in the mean free path and the IGM temperature.  As
with many of the other $z \sim 6$ measurements, improved models of the
transition era -- in which the IGM is filled with ionized gas, yet the
mean free path still has sizable spatial fluctuations -- will help
with unravelling the precise implications of the dark gap
distributions and their redshift evolution.

\subsection{Quasar near-zones}
\label{sec:prox}

As already discussed, at $z \gtrsim 6$ the \Lya forest is almost
completely saturated.  Inferring the ionization state (and other
properties) of the $z \gtrsim 6$ IGM from typical regions of the \Lya
forest is therefore challenging. However, gas in spectral regions
close to the background quasar itself, in the proximity or
``near-zone'', are exposed to ionizing radiation from the nearby
quasar as well as the UVB.  These regions show some transmission
through the \Lya forest, and can hence be used to study the $z \gtrsim
6$ IGM.

The simplest measurable property is the size of these zones.
Observationally, this is usually defined as the distance over which
the continuum normalised transmission first drops below some
threshold; in \citet{Fan2006} the threshold was chosen to be $F =
0.1$, after smoothing each spectrum to $20 \Ang$ spectral resolution.
Early work interpreted the sizes of these zones as indicating the
radii of ionized bubbles expanding around the quasars into a largely
neutral IGM \citep{WyitheLoeb2004,Wyithe2005}. However, subsequent
work showed that the transmission profile may instead reflect only the
classical proximity effect: the transmission will fall below the
chosen threshold in a highly ionized IGM when the combined
photo-ionization rate from the quasar and background galaxies is small
enough
\citep{BoltonHaehnelt2007nz,BoltonHaehnelt2007lyb,Maselli2007,Maselli2009,Lidz2007}. This
means that the observed size is an unreliable indicator of the
position of the ionization front around the quasar; this is
essentially the usual problem that the \Lya optical depth is large
even for highly ionized gas at $z \sim 6$ (see Eq. \ref{eq:evalGP}).

Nevertheless, the size of the near-zones does evolve with
redshift near $z \sim 6$ \citep{Fan2006}.  \citet{Carilli2010} fit a
linear function to the near-zone size as a function of redshift from
$z=5.7$ to $z=6.4$, and find that the proper size -- after correcting for
differences in quasar luminosities -- drops by over a factor of two across
this narrow redshift interval. This evolution may be driven by rapid
evolution in the ionizing photon mean free path just after
reionization completes \citep{Wyithe2008}. However, the size evolution
may also partly reflect the presence of remaining diffuse neutral gas
in the IGM, and so the interpretation of the observed evolution is
still unclear.

In another approach, \citet{Mesinger2004}, \citet{MesingerHaiman2007},
and \citet{Schroeder2013} have compared the near-zone transmission in
both \Lya and \Lyb to mock spectra. The latter study compares mock
spectra with three quasar spectra at $z=6.2, 6.3$, and $6.4$ and
argues that each spectrum prefers the presence of damping wing
absorption blueward of the \Lya line, arising from the natural
broadening of the transition.  Crucially, this damping wing is
prominent only if the gas is significantly neutral
\citep{MiraldaEscude1998}.  A potential source of confusion, however,
is from high column density absorbers -- damped \Lya absorbers (DLAs)
-- which also have prominent damping wings. However, the damping wing
from a DLA has a different shape compared to extended, diffuse neutral
gas in the IGM, so these two possibilities may be distinguished, at
least in principle \citep{MiraldaEscude1998}.

In practice, however, it can be difficult to distinguish damping wings
from possible diffuse neutral gas and DLAs over a limited stretch of
spectrum.  In addition, multiple neutral regions may each contribute
through their damping wings to the absorption in a given pixel and
impact the form of the damping wing profile from the diffuse IGM
\citep{MesingerFurlanetto2008,McQuinn2008,MalloyLidz2015}.  Based on
the spectra alone, \citet{Schroeder2013} cannot rule out the
possibility that the putative damping wing comes from a DLA. However, DLAs of the required column density
should be rare and these authors find that the data prefer incomplete reionization at $z
\sim 6$ and a neutral fraction of $\avg{x_{\rm HI}} \geq 0.1$ at
$95\%$ confidence.

\begin{figure}
\begin{center}
\includegraphics[width=\columnwidth]{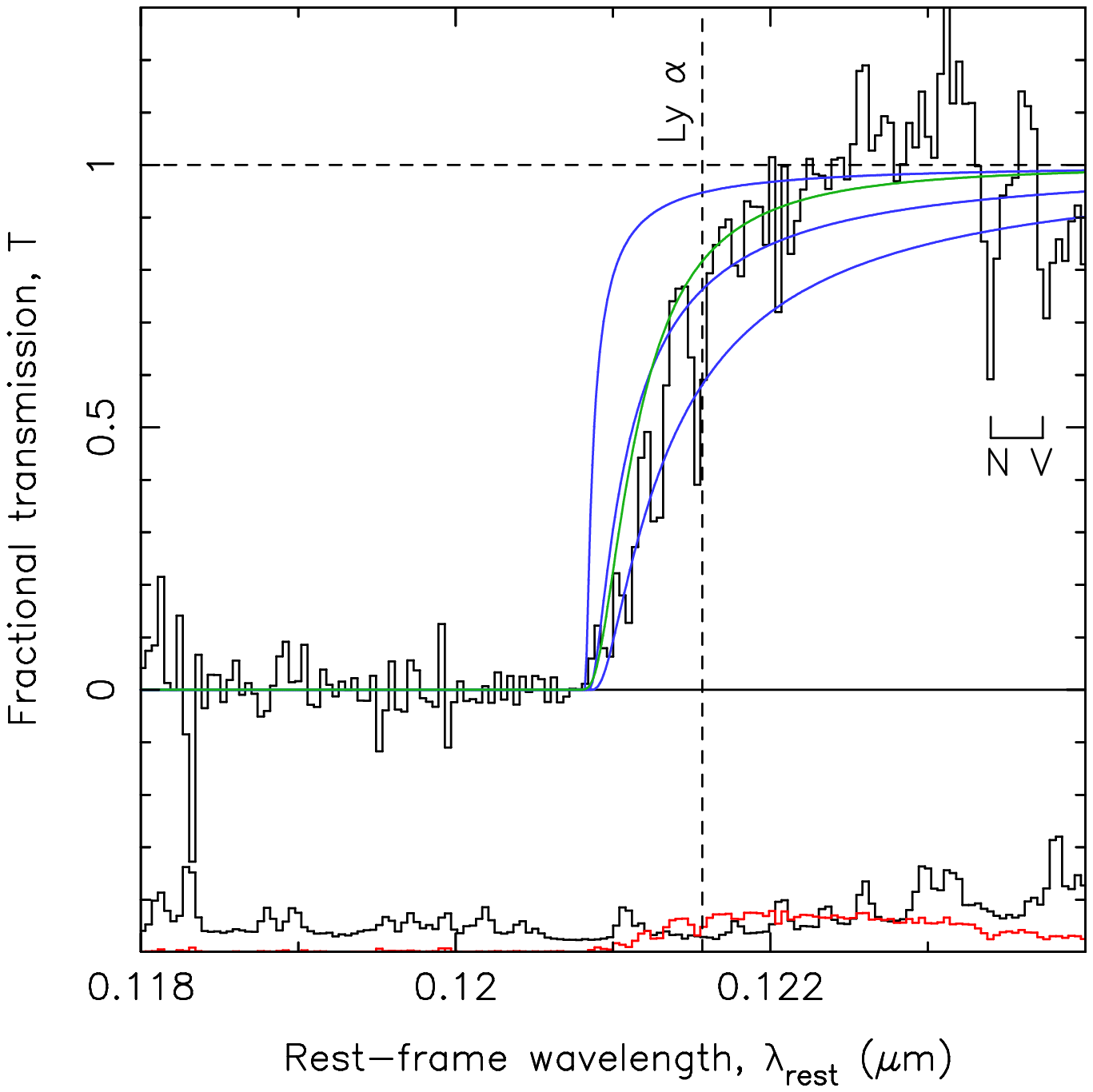}
\vspace{-1cm}
\caption{Possible damping wing feature in the spectrum of a $z=7.1$
  quasar. The black binned data shows an estimate of the transmission
  near the systemic redshift (dashed line) of the quasar ULAS
  J1120+0641.  The blue lines show damping wing absorption models
  (assuming a fully transparent ionized zone around the quasar for
  simplicity), and uniform neutral fractions of $\avg{x_{\rm HI}} =
  0.1, 0.5$, and $1.0$, with the more neutral models giving more
  absorption. The ionized zone is assumed to end abruptly, $2.2$ Mpc
  in front of the quasar. The green solid line shows an alternate fit
  in which the IGM is highly ionized and the damping wing is instead
  sourced by a DLA of column density $N_{\rm HI} = 4 \times 10^{20}$
  cm$^{-2}$, situated $2.6$ Mpc in front of the quasar. {\it
    Reproduced from Figure 4 of \citet{Mortlock2011} by permission of the authors and Macmillan
    Publishers Ltd.}}
\label{fig:dwing_mortlock}
\end{center}
\end{figure}

Perhaps the most intriguing near-zone result comes from the spectrum
of the highest-redshift quasar presently known, at $z=7.1$, which
shows possible evidence for damping wing absorption on the red side of
its \Lya line \citep{Mortlock2011}.  The spectrum close to the \Lya
emission line, along with toy damping wing models for a partly neutral
IGM and a DLA, are shown in Fig. \ref{fig:dwing_mortlock}.  This
comparison suggests that the damping wing could be sourced by
significantly neutral material, although the fully neutral model
produces too strong a wing.  \citet{Bolton2011} performed a detailed
comparison of the transmission profile with simulated spectra, and
find that either a neutral fraction of $\avg{x_{\rm HI}} \gtrsim 0.1$
is required, or a highly ionized IGM can be reconciled with the data
if a DLA lies within $\sim 5$ proper Mpc of the quasar (see also
\citealt{Keating2015}).  There are two challenges for the DLA scenario
however. First, a DLA of the required column density should be rare:
in their model, \citet{Bolton2011} find a $5\%$ probability for a DLA
of the requisite properties.  A second argument against a DLA is that
there is no detectable metal line absorption at the same redshift as
the damping wing feature \citep{Simcoe2012}. Quantitatively, this
study derives an upper limit on the metal abundance of $\leq 10^{-4}$
times the solar abundance for a dense absorber, although the feature
could nevertheless be due to a very low-metallicity proto-galaxy.

An alternative use of these proximity or near-zones is measuring the
temperature of the IGM at $z\simeq 6$
\citep{Bolton2010,Bolton2012,Padmanabhan2014}; the long cooling time
for the low-density gas in the IGM allows the gas temperature to be
used as a probe of the timing of reionization
\citep{MiraldaRees1994,Theuns2002b,HuiHaiman2003}.  High-redshift
temperature measurements close to the epoch of reionization are
therefore desirable; this limits the time available for gas to cool
following reionization and also bypasses the expected heating from
\HeII reionization at $z<4$.  In an analysis of gas temperatures
inferred from the line widths in near-zones observed in seven high-resolution ($R\sim 40\,000$) quasar spectra at $5.8<z<6.4$,
\cite{Raskutti2012} concluded these data were consistent with
reionization completing at $z>6.5$ at 95\% confidence.  However, this
inference is model dependent, and relies on assumptions for the
typical spectrum of ionizing sources during reionization as well as
the amount of photo-heating by the quasar itself.

Finally, note that an important systematic concern in all these
analyses is fitting the unabsorbed quasar continuum close to the \Lya
line. \citet{Simcoe2012} investigate this issue, considering four
different composite spectra derived from various lower-redshift
observations, and also perform a principle component analysis in an
effort to extrapolate the continuum estimate from the red side of
Ly$\alpha$. They find that the damping wing fits are fairly stable
across the range of continuum fits considered.  Nevertheless, it may
be helpful to further explore the range of possible continuum shapes
close to \Lya in lower-redshift spectra; these spectra can serve as a
control sample from an epoch when the IGM is certainly ionized
\citep{KramerHaiman2009,BosmanBecker2015}.

\subsection{Consistency with other probes of reionization}
\label{sec:consistency}

Having summarized the implications of current quasar absorption line
observations for our understanding of the EoR, we now consider these
studies in the context of a wide range of complementary
multi-wavelength EoR observations.  We start with a brief
description of existing reionization probes, and then turn to consider
current constraints from these observations.

\begin{itemize}
\item {\bf CMB:} As CMB photons propagate from the surface of last
  scattering, they may scatter off the free electrons that are (again)
  prevalent during and after reionization. This has the effect of
  damping the primary temperature anisotropies and generating
  large-scale polarization anisotropies
  \citep{Zaldarriaga1997,HuWhite1997}, while also producing secondary
  anisotropies through the kinetic Sunyaev-Zel'dovich (kSZ) effect
  \citep{GruzinovHu1998,Zahn2005,McQuinn2005}. The distinctive large
  scale polarization signal depends on the total probability that a
  CMB photon Thomson scatters off a free electron along the line of
  sight.  This is quantified by the electron scattering optical depth,
  $\tau_e$; this measurement therefore constrains an integral over the
  entire reionization history.  The patchy kSZ effect results when CMB
  photons scatter off free electrons in ionized regions during the EoR
  and receive a redshift or blueshift owing to the peculiar velocity
  of these regions.  The small-scale CMB fluctuations that are induced
  arise in part because of spatial variations in the ionization field
  \citep{Aghanim1996,GruzinovHu1998}.

\item {\bf UV luminosity functions:} Measurements using the Wide
  Field Camera 3 (WFC3) on board the {\em Hubble Space Telescope}
  (\emph{HST}) have found large populations of high-redshift galaxy
  candidates -- some of which are spectroscopically confirmed -- using
  the Lyman-break technique (see also \S\ref{sec:gals}).
  Ground-based surveys have also played an important role at the
    bright end of the luminosity function \citep{Bowler2014}.  These
  observations have allowed measurements of the UV luminosity function
  of Lyman-break galaxies (LBGs) at rest-frame wavelength of $1500
  \Ang$ out to $z \sim 8$
  \citep{Oesch2013,McLure2013,Finkelstein2014,Ishigaki2015}.  There
  are also now a handful of candidates out to $z \sim 9$--$10$
  \citep{Bouwens2014,McLeod2015}.  These measurements allow the
  contribution of observed galaxies to the overall ionizing photon
  budget to be assessed
  \citep{DuncanConselice2015,Robertson2015,Bouwens2015}, modulo
  uncertainties in the escape fraction, the galaxy UV spectral energy
  distributions, and the contribution from faint, unobserved sources
  (see discussion in \S\ref{sec:gals}).

\item {\bf Lyman-$\alpha$ emitter surveys:} Another highly successful
  approach for finding high-redshift galaxies uses a narrow-band
  selection technique to target objects with prominent \Lya emission
  lines (see \citealt{Dijkstra2014rv} and references therein). The
  visibility of these \Lya emitters (LAEs) is impacted by the damping wing
  arising from neutral gas in the IGM \citep{MiraldaEscude1998}, and so
  these surveys are sensitive to the reionization history.
  Consequently, the abundance of observable LAEs will drop and their clustering
  will increase as one probes deeper into the EoR
  \citep{Furlanetto2006,McQuinn2007lae,MesingerFurlanetto2008lae}.

\item {\bf GRB optical afterglows:} The optical afterglow spectra of
  gamma-ray bursts (GRBs) are potentially useful in searching for the
  damping wing redward of the \Lya emission line
  \citep{Totani2006,Totani2014,Chornock2013}. These sources are
  extremely luminous (for a short time) and so can be detected at high
  redshift.  Compared to quasars, GRB optical afterglows also have a
  simpler unabsorbed continuum spectrum and are more likely to occur
  in typical ionized regions (which are smaller than the ionized
  bubbles around quasars). On the other hand, GRB afterglow spectra
  typically show high-column density DLA absorption from neutral
  hydrogen in the host galaxy. The absorption from the GRB host can
  overwhelm any damping wing signature from diffuse neutral gas in the
  IGM.  However, an afterglow spectrum without prominent host absorption will
  allow a detection of --- or sharper upper limits on the presence of
  -- diffuse neutral gas in the IGM \citep{McQuinn2008}.

\item {\bf Redshifted 21 cm line:} Perhaps the most promising way of
  studying reionization is to detect redshifted 21 cm emission from
  neutral gas in the IGM during reionization
  \citep{Madau1997,Furlanetto2006rv}. First generation surveys aiming
  to detect this signal have started to place upper limits on the
  amplitude of redshifted 21 cm fluctuations
  \citep{Paciga2013,Parsons2014,Dillon2014}. These limits do not yet
  place interesting constraints on the ionization history, however,
  and are therefore not discusssed further here.  We nevertheless
  mention this probe in passing, since it may ultimately provide the
  most powerful approach for studying reionization.

\end{itemize}

\begin{figure}
\begin{center}
\includegraphics[width=\columnwidth]{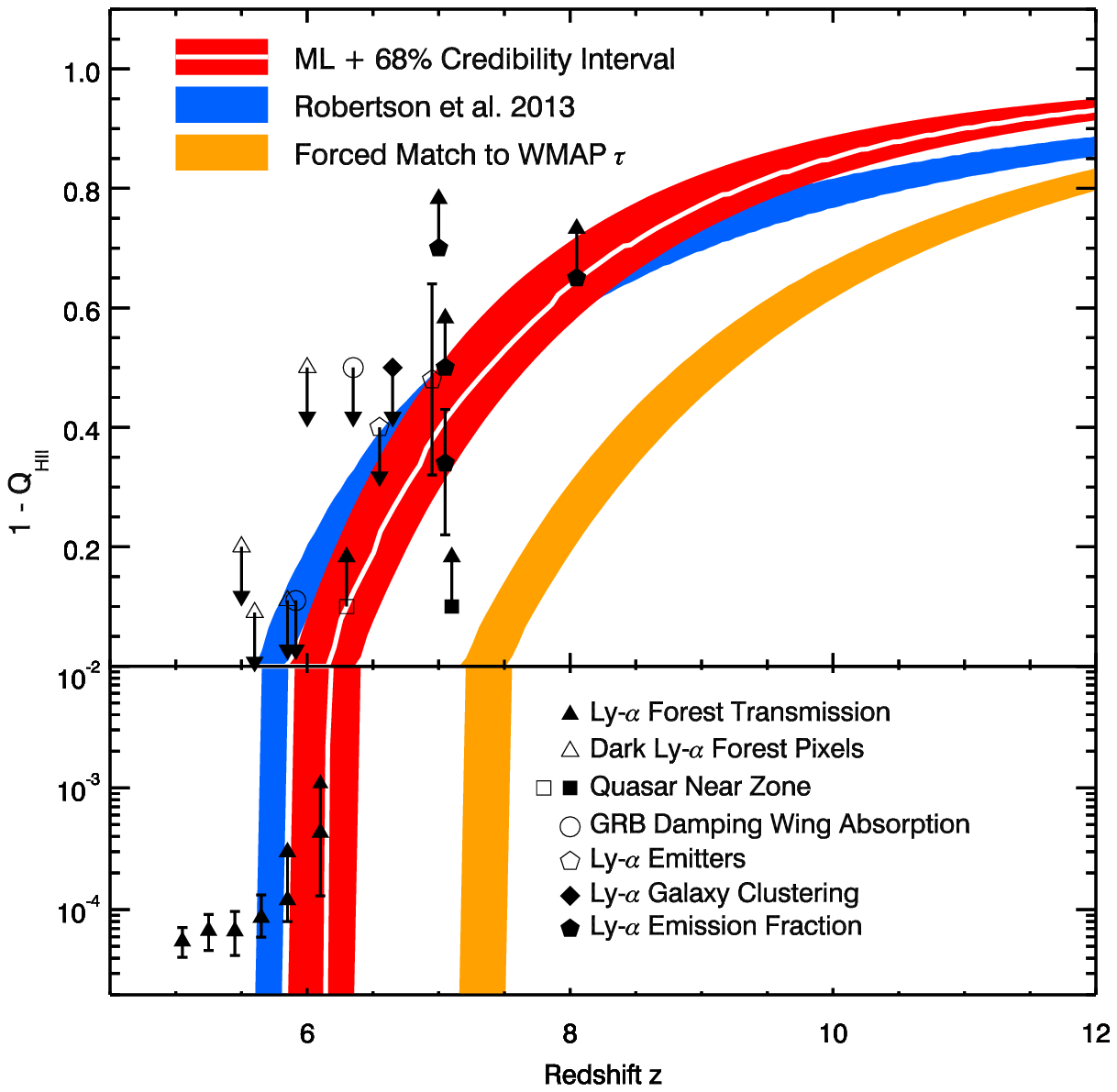}
\vspace{-0.8cm}
\caption{A summary of current multi-probe reionization
  constraints. The filling factor of neutral hydrogen is displayed as
  a function of redshift.  The constraints shown in the legend are
  from the \Lya forest transmission \citep{Fan2006}, dark \Lya forest
  pixels \citep{McGreer2015}, quasar near-zones
  \citep{Mortlock2011,Bolton2011,Schroeder2013}, GRB damping wing
  absorption \citep{Chornock2013}, \Lya emitters
  \citep{Ota2008,Ouchi2010}, \Lya galaxy clustering
  \citep{McQuinn2007lae,Ouchi2010} and the \Lya emission fraction
  \citep{Schenker2014,Pentericci2014}.  The white line and red band
  show a reionization history consistent with the observed star
  formation rate density and the \emph{Planck} electron scattering optical
  depth \citep{PlanckXIII2015}.  The orange band shows a model
  consistent with the earlier \emph{WMAP}-9 constraint. {\it Reproduced from
    Figure 3 of \citet{Robertson2015} by permission of the authors and the AAS.}}
\label{fig:xi_summary}
\end{center}
\end{figure}

\noindent
A selection of current constraints on reionization are summarized in
Fig. \ref{fig:xi_summary}, from \citet{Robertson2015}.  We first
consider constraints on $\tau_e$ from the large scale polarization CMB
power spectrum measurements. {\em Wilkinson Microwave Anisostropy
  Probe} (\emph{WMAP}) nine-year E-mode polarization data, combined with
\emph{Planck} temperature anisotropy data, give $\tau_e =
0.089^{+0.012}_{-0.014}$ (1-$\sigma$ error bars,
\citealt{PlanckXVI2013,Bennett2013}); this optical depth corresponds
to an (instantaneous) reionization redshift of $z_{\rm r} = 11.1 \pm
1.1$.  Recent large-scale polarization data from the 2015 \emph{Planck} data
release, however, suggest a smaller optical depth of $\tau_e = 0.066
\pm 0.016$, when \emph{Planck} polarization data are combined with
temperature anisotropy and CMB lensing measurements from \emph{Planck}
\citep{PlanckXIII2015}.  The optical depth\footnote{Note this
    corresponds to the ``TT+lowP+lensing'' constraint in Table 4
    reported by \cite{PlanckXIII2015}.  The optical depth varies somewhat with the
    precise data included, e.g. the ``TT+lowP'' constraint from only temperature and
    polarization anisotropy data (i.e., without CMB lensing) gives $\tau_e = 0.078 \pm 0.019$. However, \emph{all combinations}
    of the recent \emph{Planck} measurements prefer a lower optical depth than the \emph{WMAP} nine year results quoted above.
    This is mostly driven by improvements
  in cleaning polarized foreground emission; these improvements are
  enabled by the new \emph{Planck} 353 GHz polarization maps.} preferred by
\emph{Planck} corresponds to an instantaneous reionization redshift of
$z_{\rm r} = 8.8^{+1.7}_{-1.4}$. In practice, this measurement
translates into an end-point for reionization which may be close to
$z\simeq 6$.

Second, we consider the patchy kSZ effect. Studies using the
  South Pole Telescope (SPT) placed upper limits on the amplitude of
CMB fluctuations induced by the patchy kSZ effect at multipole moments
of $\ell \sim 3,000$ (\citealt{Zahn2012,George2015}, see also
\citealt{Mesinger2012}).  A grid of patchy ksZ effect models have been
used to place an upper limit on the duration of reionization of
$\Delta z \leq 7.9$ at 95\% confidence \citep{Zahn2012}. The duration
here is defined as the redshift interval over which the ionization
fraction transitions from $\avg{x_i} = 0.20$ to $\avg{x_i} = 0.99$.
Note, however, this limit is dependent on separation from other
contributions to the anisotropies, such as the thermal SZ (tSZ) effect
and the Cosmic Infrared Background (CIB), using their differing
frequency and angular scale dependence.  Here we have quoted the most
conservative constraints that allow a large (anti-)correlation between
the tSZ and CIB: in the absence of any correlation, the 95\%
confidence limit shrinks to $\Delta z \leq 4.4$.  A more recent
analysis by \cite{George2015} finds a similar value of $\Delta z \leq
5.4$ at 95\% confidence.

The implications of the UV luminosity functions of LBGs for reionization
are discussed in \S\ref{sec:gals}. Here we simply remark that
although the details depend on the assumed escape fraction, ionizing
spectrum, and faint-end extrapolation, these observations are
compatible with reionization completing late, at $z \leq 6$ or so.
The red band in Fig.~\ref{fig:xi_summary} displays a reionization model
consistent with UV luminosity functions extrapolated to absolute
magnitudes of $M_{\rm UV}\simeq -13$ and a constant escape fraction
$f_{\rm esc}=0.2$ \citep{Robertson2015}.  This is also in accord with
the recent \emph{Planck} optical depth measurements \citep{PlanckXIII2015}.

Results from a selection of LAE surveys are also shown in
Fig. \ref{fig:xi_summary}. Recent studies of the abundance of LAEs
detect a relatively small drop from $z=5.7$ to $z=6.6$, while these
surveys are thus far finding very few galaxies at
$z=7.3$.\footnote{These redshifts ($z=5.7, 6.6, 7.3$) correspond to
  narrow gaps between bright night sky lines, where LAEs can be found
  efficiently from the ground using the narrow band technique.}  Based
on the lack of LAE candidates in the Subaru Deep Field at $z=7.3$,
\citet{Konno2014} argue that the neutral filling factor is
$\avg{x_{\rm HI}} = 0.3-0.8$ at $z=7.3$. From the $z=6.6$ LAE
abundance, \citet{Ouchi2010} place an upper limit of $\avg{x_{\rm HI}}
\leq 0.4$, and from the lack of strong LAE clustering these authors
conclude that $\avg{x_{\rm HI}} \leq 0.5$. Finally, from measuring the
redshift evolution of the \Lya fraction -- the fraction of LBGs
that emit appreciably in \Lya -- \citet{Schenker2014} find
$\avg{x_{\rm HI}} = 0.34^{+0.09}_{-0.12}$ at $z=7$ and $\avg{x_{\rm
    HI}} > 0.65$ at $z=8$ (1-$\sigma$).  Note, however, these
inferences are model dependent; see e.g.
\citet{BoltonHaehnelt2013,Dijkstra2014,TaylorLidz2014,Pentericci2014,Choudhury2014,Mesinger2015}
for further discussion regarding these results.

Finally, current high-redshift GRB optical afterglow spectra have
mostly been limited by the presence of strong host DLA absorption, as
we alluded to previously. Nevertheless, some limits are possible.  For
example, \citet{Totani2006} find $\avg{x_{\rm HI}} < 0.60$ at 95\%
confidence at $z=6.3$, while \citet{Chornock2013} find a 2$\sigma$
upper limit of $\avg{x_{\rm HI}}<0.11$ at $z=5.91$.

In summary, when compared to the constraints from quasar absorption
lines discussed earlier in this section and displayed in
Fig.~\ref{fig:xi_summary}
\citep{Fan2006,Bolton2011,Mortlock2011,Schroeder2013,McGreer2015}, at
present the various probes appear broadly consistent with each
other. The current data point to a scenario where the completion
redshift (when the filling factor of ionized hydrogen reaches unity)
is relatively late, between $5.5 \leq z \leq 7$.  However, as
discussed there remain many significant uncertainties associated with
all these measurements -- the challenge is therefore to make a more
precise and definitive statement.  As we now briefly discuss, in the
next decade the prospects for further progress toward this goal are
promising.

\section{Conclusions and future prospects} 
\label{sec:conclude}

Quasar absorption line studies yield a rich body of information
describing the ionization, thermal, and chemical enrichment history of
the intergalactic gas, allowing us to study the evolution of most of
the baryons in the Universe all the way out to -- and perhaps into --
the EoR.  The properties of the intergalactic gas are in turn
determined by the underlying cosmic web of structure formation -- as
specified, for example, according to the currently favored
inflationary, cold dark matter cosmological model with dark energy --
along with the properties of galaxies and AGN, which strongly
influence the gas in their surroundings.  The combination of IGM
measurements with direct censuses of galaxies and AGN then provide a
powerful probe of early galaxy and structure formation.

\begin{figure*}
\begin{center}
\includegraphics[width=0.45\textwidth]{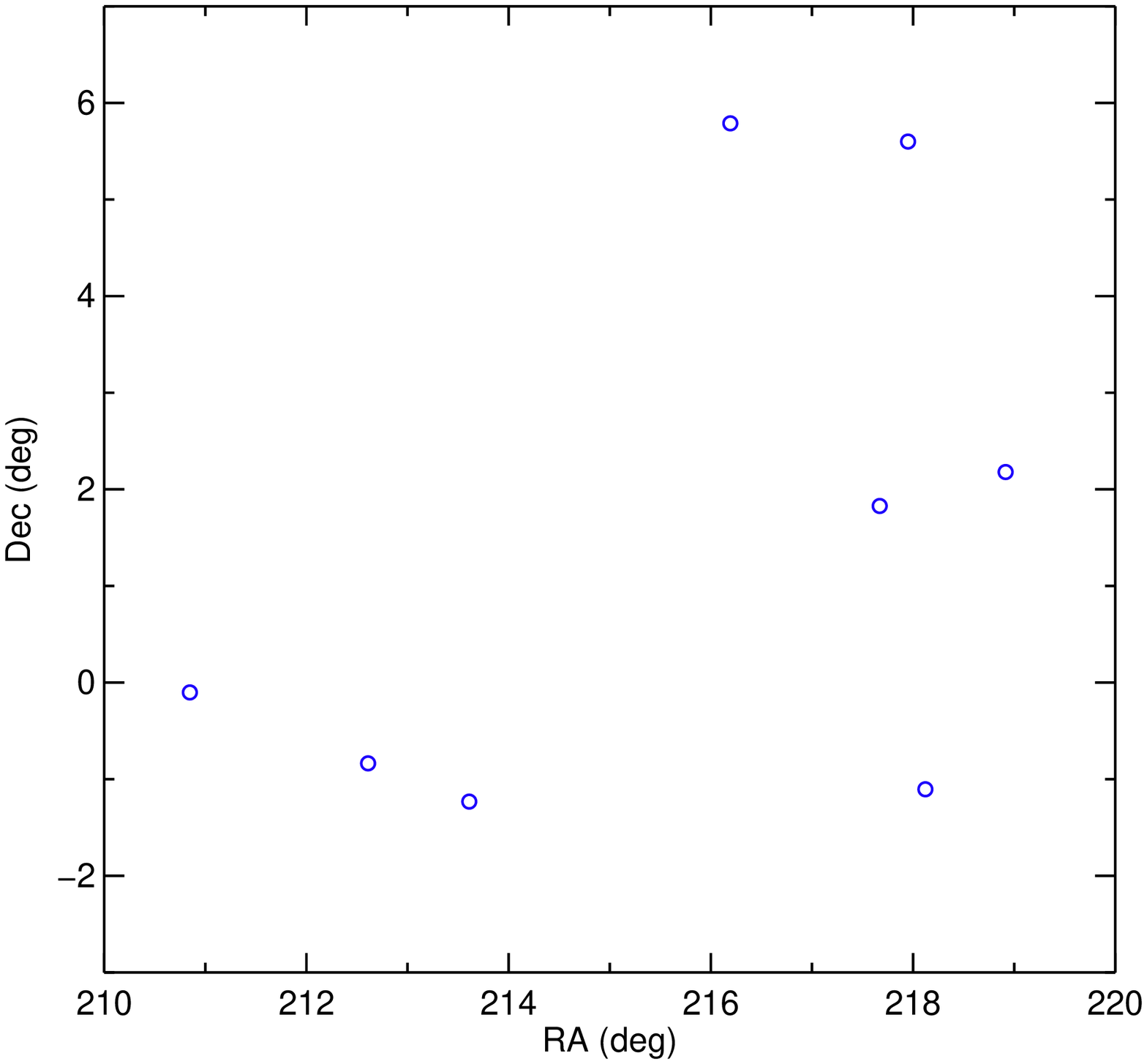}
\includegraphics[width=0.45\textwidth]{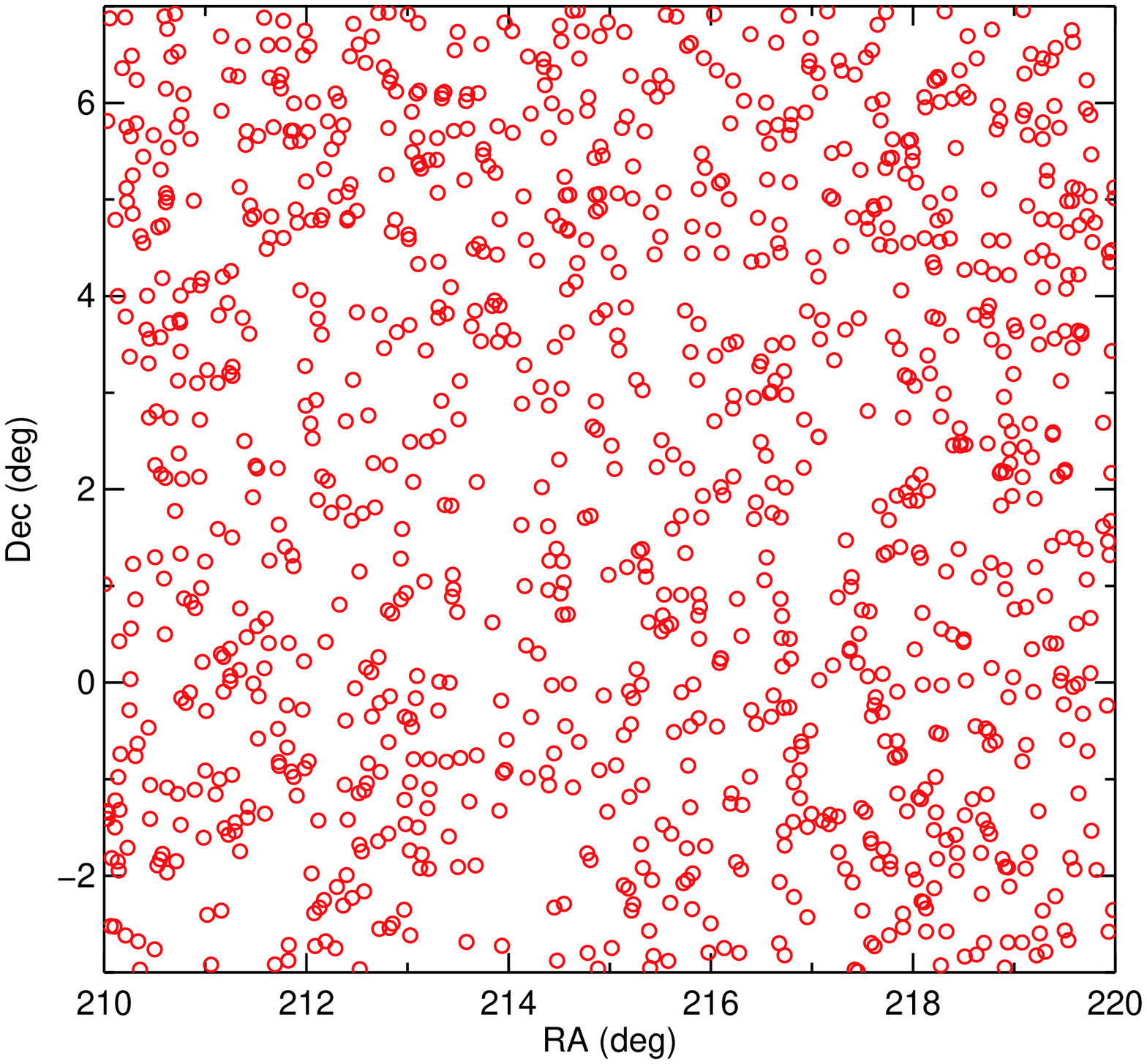}
\vspace{-0.4cm}
\caption{Illustration of the increased density of quasar targets for
  spectroscopic follow-up made accessible by $30$ meter-class
  telescopes. The left-hand panel shows the quasars in a 100 deg$^2$
  patch of the sky surveyed by the Baryon Oscillation
    Spectroscopic Survey (BOSS) out to $m_r < 18$.  These represent
  targets for which high-quality data can be obtained in $\sim$1 hour
  of observing using high-resolution spectrographs on current $8$
  meter telescopes.  For contrast, the right-hand panel shows the
  quasars -- in the same patch of sky -- that may be readily observed
  using a high-resolution spectrograph on a $30$ meter telescope ($m_r
  < 21$).}
\label{fig:bigt_sampling}
\end{center}
\end{figure*}

In the next 5-10 years, there are excellent prospects for
improving our understanding of the EoR using quasar absorption lines.
First, we emphasize that several independent observations, summarized
in Fig.~\ref{fig:xi_summary}, are now converging towards reionization
histories that complete sometime between $5.5 \leq z \leq 7$; these
are redshifts at which significant amounts of quasar absorption line
data already exist. This further motivates studies aimed at
understanding reionization's precise signatures -- and that of
reionization's immediate aftermath -- in absorption line
spectra. Renewed efforts to model the challenging end-phases of the
EoR, in conjunction with novel data analysis techniques, may therefore
lead to breakthroughs in our understanding of the EoR using only {\em
  currently available data.}

Still more compelling, though, will be the advances enabled by future
data sets.  Current optical and near-infrared wide-area photometric
surveys such as those from Panoramic Survey Telescope and Rapid
  Response System (Pan-STARRS), the Dark Energy Survey (DES),
and the Visible and Infrared Survey Telescope for Astronomy
(VISTA) are already increasing the number of known quasars at $z \sim
6-7$ \citep[e.g.,][]{venemans2013,venemans2015,banados2014,Reed2015}.
Over the next several years these surveys should roughly double the
number of known quasars at $z \sim 6$ and provide the first
significant samples at $z \geq 7$.  These quasars will provide greatly
improved statistics for metal line, \Lya forest, and quasar near-zones
studies at $z \geq 6$.  For example, they should reveal whether the
possible red damping-wing feature observed by \citet{Mortlock2011} at
$z \sim 7$ is ubiquitous amongst quasars near this redshift, and may
uncover evidence of chemical enrichment from metal-free stars.
Further dramatic increases in the number of known high-redshift
quasars will come from deep surveys conducted by the Large
  Synoptic Survey Telescope (LSST), \emph{Euclid}, and the \emph{Wide-Field
  Infrared Survey Telescope} (WFIRST).

In the next decade, there will be further progress from high-resolution spectrographs on $30$ meter-class telescopes, such as the
Thirty Meter Telescope (TMT), the Giant Magellan
  Telescope (GMT), and the European Extremely Large Telescope
(E-ELT). The enhanced collecting area of these telescopes will allow
significant improvements in the spectroscopy of high-redshift
quasars. First, it will enable spectroscopic follow-up of much fainter
quasars than is currently feasible in reasonable integration
times. Since the quasar luminosity function is quite steep, this
translates into a significant boost in the density of quasar
sightlines that may be probed spectroscopically, as illustrated in
Fig.~\ref{fig:bigt_sampling}.  This will improve statistics, and
especially provide large samples of close quasar pairs at high
spectral resolution; this can be used to measure, among other
quantities, the pressure or Jeans smoothing scale, which is sensitive
to the timing of reionization \citep[e.g.][]{Rorai2013}.

\begin{figure}
\includegraphics[width=0.49\textwidth]{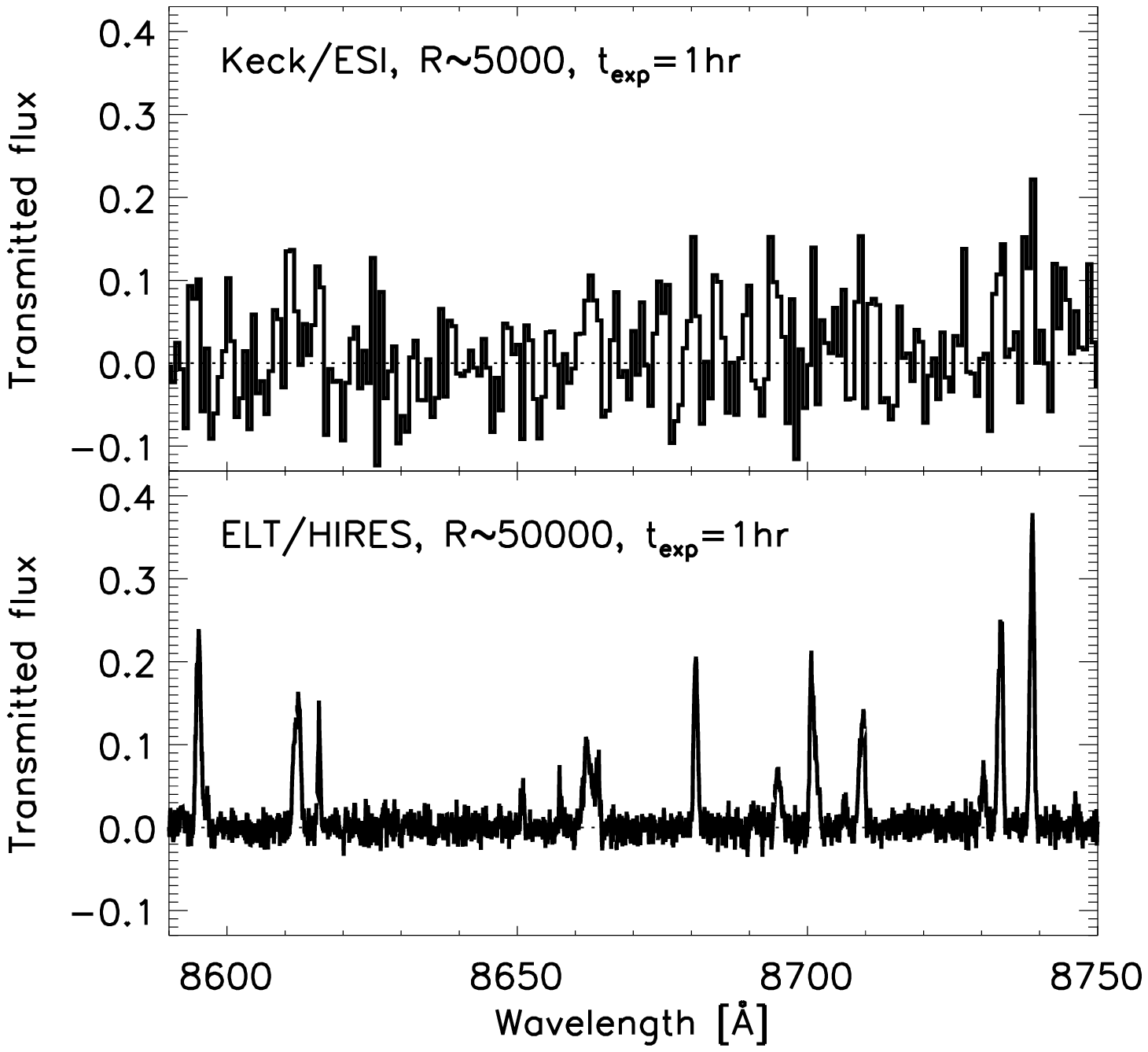}
\vspace{-0.8cm}
\caption{An example of the sensitivity gains enabled by future 30-40
  meter class telescopes.  The top panel shows a simulated region of
  \Lya forest at $z \simeq 6.1$ in the spectrum of a moderate
  luminosity quasar with magnitude $z_{\rm AB} = 20.5$ as it would be
  observed in roughly a one-hour exposure with the 10 meter
  Keck/Echellete Spectrograph and Imager (ESI).  The bottom
  panels shows the same spectrum obtained with a high-resolution
  spectrograph on a 39 meter telescope.  Note the narrow transmission
  peaks from ionized regions which are now clearly detected and
  resolved.}
\label{fig:elt_quasar}
\end{figure}

In addition to the higher density of observable sightlines, it will be
practical to obtain higher resolution and signal-to-noise ratio
spectra in relatively short exposure times. This is illustrated in
Fig.~\ref{fig:elt_quasar}, which shows mock spectra of the \Lya forest
at $z=6.1$, comparing what is presently achievable (at a spectral
resolution of $R \sim 5\,000$, e.g. \citealt{White2003,Fan2006}) with
what will be accessible to a high resolution ($R \sim 50\,000$)
spectrograph on a $30$ meter class telescope. This improved
spectroscopy will make it easier to discern whether ``fully absorbed''
regions in the forest are truly saturated, facilitate the detection of
weak metal absorption lines, and will help resolve \Lya line-widths,
facilitating measurements of the temperature of the IGM at $z \geq 5$
(e.g.\citealt{Becker2011,Trac2008,FurlanettoOh2009,LidzMalloy2014}),
including temperature measurements in the quasar near-zones at $z \geq
6$ \citep{Bolton2012,Raskutti2012}.

In conjunction with improved quasar absorption line studies, we expect
an enhanced interplay with other reionization probes in the near
future.  The \emph{James Webb Space Telescope} (\emph{JWST}) is scheduled to
launch in 2018, and will allow detailed studies of early galaxy
populations in the infrared, while the Atacama Large Millimeter
  Array (ALMA) may detect the same galaxies at sub-millimeter
wavelengths through dust emission, and molecular and atomic
fine-structure emission lines
\citep[e.g.][]{Maiolino2015,Capak2015,Willott2015,Watson2015}. The
Subaru Hyper Suprime-Cam will map-out the positions of sizable
populations of LAEs out to $z=7.3$ over wide regions of the sky,
enabling improved measurements of the abundance and clustering of
these galaxies; these observations will in turn inform our
understanding of the ionization state of the surrounding IGM. Finally,
redshifted 21 cm surveys will provide a direct probe of the neutral
hydrogen in the IGM during the EoR, and potentially earlier phases of
cosmic structure formation
\citep{Furlanetto2006rv,Paciga2013,Yatawatta2013,Parsons2014,Dillon2014}. Insights
gained from quasar absorption lines can help plan and optimize the
observing strategy for many of the other future surveys. In addition,
absorption line studies will continue to play an important role in
piecing together a consistent picture of the EoR in conjunction with
the full suite of upcoming observations. For example, quasar near-zone
measurements, LAE abundance and clustering observations, GRB optical
afterglow spectra, and redshifted 21 cm surveys may all provide
independent measurements of the volume-averaged neutral fraction over
overlapping redshift ranges.  Metal absorption lines, combined with
direct galaxy observations with \emph{JWST} and ALMA will further help to
reveal the nature of the galaxies responsible for reionization.  In
summary, the prospects for greatly expanding our understanding of the
reionization epoch are excellent over the next several years.

\begin{acknowledgements}
We thank Kristian Finlator, Emma Ryan-Weber, Rob Simcoe, and the anonymous referee for helpful comments.  JSB acknowledges the support of a Royal Society University Research
Fellowship. AL acknowledges support from NASA grant NNX12AC97G and
from the NSF through grant AST-1109156.
\end{acknowledgements}

\bibliographystyle{apj}
\bibliography{references}

\end{document}